\documentclass{emulateapj}
\usepackage{graphicx}
\usepackage{longtable}
\usepackage{amssymb}
\usepackage{epstopdf}
\begin{document}
\hyphenation{Form-al-de-hyde dis-eq-uil-ib-ri-um ap-prox-i-mated}

\pagenumbering{arabic}

\shorttitle{Planetary Disequilibrium Chemistry}
\shortauthors{Zahnle \& Marley}

\title{Methane, Carbon Monoxide, and Ammonia in Brown Dwarfs and Self-Luminous Giant Planets}

\author{Kevin J. Zahnle}
\affil{NASA Ames Research Center, MS-245-3, Moffett Field, CA 94035; Kevin.J.Zahnle@NASA.gov}

\author{Mark S. Marley}
\affil{NASA Ames Research Center, MS-245-3, Moffett Field, CA 94035; Mark.S.Marley@NASA.gov}

\begin{abstract}

We address disequilibrum abundances of some simple molecules in the atmospheres of solar composition brown dwarfs and self-luminous extrasolar giant planets using a kinetics-based 1D atmospheric chemistry model.
Our approach is to use the full kinetics model to survey the parameter space with
effective temperatures between 500 K and 1100 K. 
In all of these worlds equilibrium chemistry favors CH$_4$ over CO in the parts of the atmosphere that can be seen from Earth,
but in most disequilibrium favors CO. 
The small surface gravity of a planet strongly discriminates against CH$_4$ when compared to an otherwise comparable brown dwarf. 
If vertical mixing is like Jupiter's, the transition from methane to CO occurs at 500 K in a planet.
Sluggish vertical mixing can raise this to 600 K;
but clouds or more vigorous vertical mixing could lower this to 400 K.
The comparable thresholds in brown dwarfs are $1100\pm100$ K. 
Ammonia is also sensitive to gravity, 
but unlike CH$_4$/CO, the NH$_3$/N$_2$ ratio is insensitive to mixing, which
makes NH$_3$ a potential proxy for gravity.
HCN may become interesting in high gravity brown dwarfs with very strong vertical mixing.
Detailed analysis of the CO-CH$_4$ reaction network reveals that the bottleneck to CO hydrogenation
goes through methanol, in partial agreement with previous work. 
Simple, easy to use quenching relations are derived by fitting to the complete chemistry of the full ensemble of models.
These relations are valid for determining CO, CH$_4$, NH$_3$, HCN, and CO$_2$ abundances
in the range of self-luminous worlds we have studied but may not apply if atmospheres are strongly heated 
at high altitudes by processes not considered here (e.g., wave breaking).
 
\end{abstract}
\section{Introduction}

\medskip

Disequilibrium chemistry has been known in Jupiter's atmosphere for several decades
\citep{Prinn1977,Bezard2002}
and has been expected and suspected in brown dwarf atmospheres from the time of their discovery 
\citep{Fegley1996, Noll1997, Saumon2000}.
The most famous disequilibrium is an overabundance of CO relative to CH$_4$.    
 This occurs in Jupiter and brown dwarfs when CO is dredged up from deep, hot layers of the atmosphere 
 more quickly than chemical reactions with ambient hydrogen can convert it to CH$_4$. 
 It is to be expected that similar processes take place in young, self-luminous extrasolar giant planets.
 But the apparent paucity of methane in the atmospheres of planets with effective temperatures comparable
 to those of methane-rich T-type brown dwarfs was underpredicted and has been met with surprise.
 
 Among field brown dwarfs methane appears---by definition---at the L to T-type transition where their near-infrared colors turn to the blue at effective temperatures near 1200 K \citep{Kirk2005}. The first directly imaged planets, however, were found to have effective temperatures below 1200 K and yet their near-infrared spectra were devoid of signs of methane in K band \citep[e.g.,][]{Barman2011a}.
The best example is HR8799c: despite an effective temperature near 1100 K,  in both low and high resolution spectra---particularly the high resolution spectrum taken by \citet{Konopacky2013}---methane has gone missing.  

Surface gravity is the defining difference between 
extrasolar giant planets and brown dwarfs, and to date is the only proven difference, although there are great hopes for metallicity. 
Field brown dwarfs (hereafter BDs) have high surface gravities ($g$ of order $10^{5}$ cm/s$^2$) and therefore very compressed scale heights.
Extrasolar giant planets (EGPs) have modest surface gravities ($g$ of order $10^3$ cm/s$^2$) and extended scale heights \citep[e.g.,][]{Burrows1997,Saumon2008}\footnote{We do not intend to wade into the nomenclature battles here. For our purposes companion objects below $\sim 13\,\rm M_J$ are planets.}.
A BD and a self-luminous EGP of the same composition and the same effective temperature will have similar optical depths as a function of temperature, at least in the absence of clouds;
i.e., the function $T(\tau)$ is roughly the same.  However, because $p \propto \tau g$,
the BD has a much higher pressure at a given optical depth than does the EGP, and a BD is much cooler than the EGP at a given pressure.
Because lower temperatures and higher pressure favor CH$_4$ in its struggle with CO,
it has been pointed out that CH$_4$ will be more easily seen and CO less easily seen at lower gravity in BDs \citep{Hubeny2007}.
Hence \citet{Barman2011a} suggested that the even lower gravity of HR8799b might be why no methane is seen in it.
We agree.
We will verify that the dependence on $g$ is strong and leads to qualitatively different outcomes for BDs and self-luminous EGPs,
a result implicit in previous work but far from fully appreciated.

\medskip
Previous studies of carbon speciation in BDs and EGPs have mostly focused on a few particular objects
 \citep{Saumon2006,Geballe2009,Barman2011a, Barman2011b, Moses2011, Line2011},
 or on hot highly irradiated EGPs \citep{Moses2011,Visscher2012}, 
or used one of several quench approximations culled from the literature 
\citep{Fegley1996,Saumon2000,Lodders2002, Hubeny2007},
or various combinations of the above \citep{Cooper2006, Visscher2011,Moses2013a,Moses2013b}. 
  The basic idea is that CH$_4$ and CO will often be seen in disequilibrium because the chemical
reactions that would enforce equilibrium don't have time enough to take place while the gas is cool \citep{Prinn1977}.
The disequilibrium composition that results is described as ``frozen-in'' or ``quenched.''
 
Quenching has been widely used to quantify discussions of CO-CH$_4$ 
and N$_2$-NH$_3$ disequilibria in a wide range
of astrophysical problems, dating at least back to \citet{Prinn1977}'s study of CO in Jupiter.
Earlier discussions of quenching can be found with respect to the N$_2$-O$_2$-NO system
that is important in thunderbolts \citep{Chameides1979}, meteor entry and rocket r{\"e}entry \citep{Park1978},
and explosions in Earth's atmosphere \citep{Zeldovich1967}. 
In the quench approximation, the disequilibrium composition that one can observe 
is approximated by the equilibrium composition when the 
relevant chemical reaction time scale $t_{\rm chem}$ equals the relevant cooling timescale, 
which if due to mixing can be written $t_{\rm mix}$ \citep{Prinn1977}.
There are many different prescriptions for defining $t_{\rm chem}$ and $t_{\rm mix}$ that we will discuss below.
The history of quench schemes for jovian planets, exoplanets, and brown dwarfs has  
been comprehensively recounted in a series of recent papers by Moses and colleagues 
\citep{Moses2010, Moses2011, Visscher2011}. 

\medskip
\medskip
Here we do something different.
We use a 1D chemical kinetics code coupled to $p$-$T$ profiles
 from a detailed 1D atmospheric structure code to compute a galaxy of 
chemical compositions in a wide range of possible brown dwarfs and cooling EGPs.
The code explicitly includes reverses of all reactions so that, in the absence of atmospheric physics,
the chemical composition would relax to equilibrium at every height.
Our strategy is to find the apparent quench points in all the models and analyze these for their systematic properties.
We then fine tune our results by using them in quench approximations.
Our objective is to describe the emergent properties of the chemical network as a whole. 
Our strategy differs from previous work that seeks to determine the one key rate-limiting step in a network of
reactions, which is then treated as the effective reaction rate for the network as a whole.

We limit the study to self-luminous cooling worlds for which insolation is not (yet) thermally important.
This includes brown dwarfs, free-floating planets, and young directly-imaged planets.
This category includes most of the exoplanets for which good data can be obtained in the present or in the near future. 
What this limitation means for carbon speciation is that we are concerned only with the conversion of CO to CH$_4$.
Unlike \citet{Line2011}, \citet{Visscher2011}, and \citet{Visscher2012},
we do not address the kinetic inhibition against oxidizing CH$_4$ to CO.
The latter is an issue in strongly irradiated planets that are warm at high altitudes where,
 if vertical mixing is fast and the temperature not too hot, CH$_4$ and its photochemical products will be overabundant
 \citep{Line2010, Miller-Ricci2011, Morley2013}.
 Methane oxidation could be an issue for BDs and cooling EGPs if the higher parts of their atmospheres
 are strongly heated by wave-breaking processes not taken into account in our radiative-convective model,
 but we do not further address this possibility here.

\medskip
Quenching in the nitrogen system has also been the subject of many studies over the years
 \citep{Abelson1966, Chameides1981, Prinn1987, Fegley1994, Fegley1996,
Saumon2006, Moses2010, Moses2011, Line2011}. 
As with methane, the underabundance  of ammonia in brown dwarfs cold
enough to favor it has been attributed to disequilibrium chemistry \citep{Saumon2006}.
The visibility of ammonia has been made the distinguishing characteristic of the Y dwarf,
the newest, coldest, and possibly last member of the stellar spectral
sequence \citep{Cushing2011}.
The NH$_3$-N$_2$-HCN system differs from CO and CH$_4$ in interesting ways
that lead to significantly different behavior. 

Finally, for completeness, we address quenching of CO$_2$, a gas that can be relatively easy to observe from 
a space-based observatory
and, because its abundance is sensitive to metallicity, can be relatively telling.  CO$_2$ is not always thought of as a
species subject to quenching (but see \citet{Prinn1987}). 
In the H$_2$-rich, UV-poor worlds that are the subject of this study, CO$_2$ quenching does take place.

\section{Overview of CO and CH$_4$} 

\begin{figure}[!htb] 
   \centering
   \includegraphics[width=0.45\textwidth]{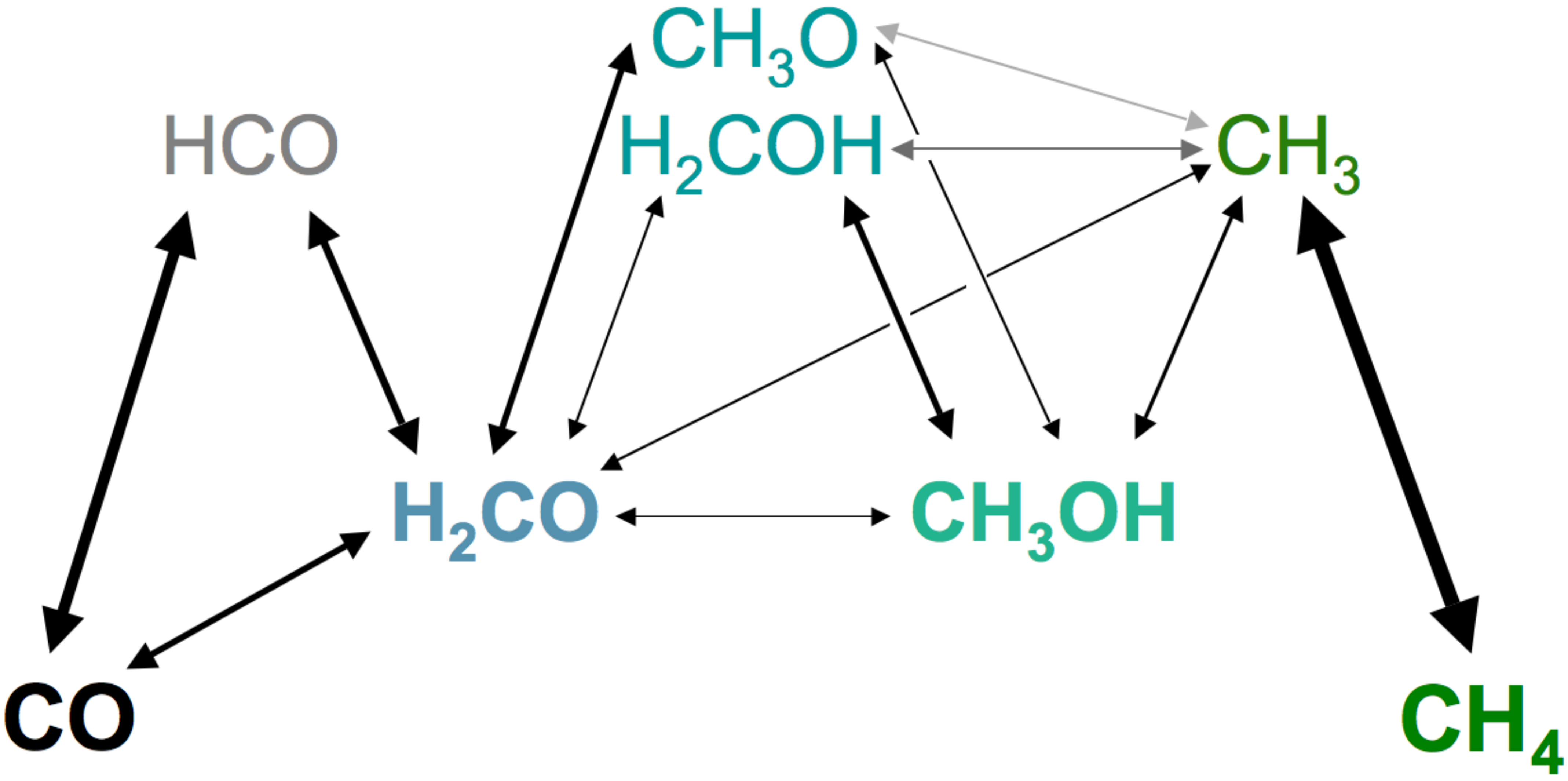} 
   \caption{\small Major chemical pathways linking CO and CH$_4$ in an H$_2$-rich atmosphere. 
     Reactions from left to right are with H$_2$ or H.
 Key intermediate molecules are formaldehyde (H$_2$CO) and methanol (CH$_3$OH).
   Other intermediates (HCO, H$_2$COH, CH$_3$O, CH$_3$) are short-lived free radicals. 
   Where the species are plotted gives a rough indication of the energetics. 
   Energy barriers correspond to breaking C-O bonds: from triple to double, from double to single, and from single to freedom.
   Relative magnitudes of reaction rates are indicated by arrow thickness for conditions near the quench point 
   in a particular model 
   (red star on Fig \ref{ensemble}) 
   that gives roughly equal amounts of CO and CH$_4$. 
   At higher temperatures and lower pressures the plot as a whole would tilt to the left, with carbon pooling in CO.
   At lower temperatures and higher pressures the plot as a whole tilts to the right and carbon pools in CH$_4$.
    The energy barriers pose greater obstacles when the gas is colder.
 }
\label{chem_figure}
\end{figure}

Figure \ref{chem_figure} illustrates the most important chemical pathways between CO and CH$_4$ in a warm H$_2$-rich atmosphere.
The path from CO to CH$_4$ climbs over three energy barriers, the first between CO and formaldehyde (H$_2$CO), the second between formaldehyde
and methanol (CH$_3$OH), and the third between methanol and methane.
The three barriers can be thought of as reducing the C$\equiv$O triple bond to a double bond, reducing the C=O double bond to a single bond,
and splitting C from O entirely. 

In photochemistry the first barrier is unimportant because atomic hydrogen is generated in
abundances that vastly exceed equilibrium \citep{Liang2003}. 
Kinetic barriers to adding H to CO to make HCO or adding H to HCO to make H$_2$CO are insignificant.  
The third barrier is unimportant in photochemistry because incident
UV radiation readily splits CH$_3$OH into CH$_3$ and OH,
  which is quickly followed by adding photochemical H to CH$_3$ to make CH$_4$. 
In photochemistry the energy to overcome the first and third barriers comes from UV photons.
  
By contrast, the middle barrier is not easily overcome by photochemistry.  
Formaldehyde is readily photolyzed but the products are either CO or HCO; i.e., regress
that leaves the C=O bond unbroken.
Meanwhile successive 3-body additions of photochemical H to H$_2$CO to make CH$_3$OH face considerable kinetic barriers,
save at high temperatures where CH$_4$ is not favored.
The historic focus of the planetary literature was therefore on reactions that go from formaldehyde to methanol. 
  \citet{Prinn1977} suggested 
  ${\rm H}_2 + {\rm H}_2{\rm CO} \rightarrow {\rm CH}_3 + {\rm OH}$,
  an ambitious reaction that jumps two hurdles at once, as the rate-limiting step, while
 Yung et al (1988) suggested that
  ${\rm H} + {\rm H}_2{\rm CO} + {\rm M} \rightarrow {\rm CH}_3{\rm O} + {\rm M}$ (where ${\rm M}$ represents a third body) would be the bottleneck. 
  Both schemes have been widely used, for Jupiter \citep{Bezard2002}, EGPs \citep{Cooper2006, Line2011}, and BDs \citep{Hubeny2007, Barman2011a}.

Without the photochemical assist, any of the three barriers could potentially block the reaction.  
Recently,  \citet{Moses2011} 
concluded that the rate-limiting reaction is the one that breaks the C-O single bond.
They report that the bottleneck is between methanol and methane,
${\rm CH}_3{\rm OH} + {\rm M} \rightarrow {\rm CH}_3 + {\rm OH} + {\rm M}$
when methane is abundant,
or goes through a free radical
 ${\rm H} + {\rm H}_2{\rm COH} \rightarrow {\rm CH}_3 + {\rm OH}$
 when methane is scarce.
 We will concur with the former but not the latter.

\section{The photochemical model}
 
Standard 1D codes simulate atmospheric chemistry by computing the gains and losses of chemical species at different altitudes while accounting for vertical transport.
Vertical transport is parameterized as a diffusive process 
with an ``eddy diffusion coefficient,'' denoted $K_{zz}$ [cm$^2$/s].
Volume mixing ratios $f_i$ of species $i$ are obtained by solving continuity
\begin{equation}
\label{eq_one}
N{\partial f_i \over \partial t} = P_i - L_i N f_i - {\partial \phi_i \over \partial z} 
\end{equation}
\noindent and force (flux)
\begin{equation}
\label{eq_two}
\phi_i = b_{ia} f_i \left( {m_a g\over kT} - {m_i g\over kT}\right) - \left( b_{ia} + K_{zz}N\right) {\partial f_i \over \partial z}
\end{equation}
\noindent equations.
In these equations $N$ is the total number density (cm$^{-3}$); $P_i - L_i N f_i$ represent chemical production and loss terms, respectively; $\phi_i$ is the upward flux; $b_{ia}$, the binary diffusion coefficient between $i$ and the background atmosphere $a$, describes true molecular diffusion; and $m_a$ and $m_i$ are the molecular masses of $a$ and $i$.
We implement molecular diffusion of a heavier gas through H$_2$
by setting $b_{ia} = 6\times 10^{19}\left(T/1400\right)^{0.75}$ cm$^{-1}$s$^{-1}$---appropriate for CO---for all the heavy species.
This is a reasonable approximation for present purposes, as our concern is with quenching at altitudes well below the homopause.
The physical meaning of $b_{ia}$ is the ratio of the relative thermal velocities of the two species to their mutual collision cross section. 
In the present circumstances, the relative thermal velocity is effectively that of H$_2$, so the only important source of variation in $b_{ia}$ 
stems from the different diameters of the molecules.  
The code has been used by \citet{Zahnle2009} to address hot Jupiters
and by \citet{Miller-Ricci2011} and \citet{Morley2013} to address H$_2$-rich hot Neptunes and super-Earths.

Temperature and pressure profiles are imported from the output of detailed
radiative-convective models that are described elsewhere \citep{Saumon2008}.
The models are cloudless and of solar metallicity.
Adding cloud opacity would make the $p$-$T$ profiles hotter and less favorable
to CH$_4$.  The cloudless model is something of a best case for methane.

Treating vertical transport as diffusion is a necessary evil in a 1-D code.
We will regard $K_{zz}$ as a mildly constrained free parameter.
In free convection, $K_{zz}$ has been estimated from mixing length theory \citep{Gierasch1985},
   \begin{equation}
   \label{Gierasch}
   K_{zz} = {1\over 3} H \left(  {R_{\rm gas} F_{\rm conv} \over \mu \rho C_p}  \right)^{1/3},
   \end{equation}
   where $R_{\rm gas}$ is the universal gas constant, $F_{\rm conv}$ the convective heat flux, $\mu$ the mean molecular weight (dimensionless), $\rho$ the density,
   and $C_p$ the heat capacity of the gas.  The mixing length is approximated by the scale height.
   An upper bound is obtained from Eq \ref{Gierasch} by equating $F_{\rm conv}$ with $\sigma T_{\rm eff}^4$,
   \begin{equation}
   \label{Gierasch2}
   K_{zz} < 2.5\times 10^{10} \left(  {T_{\rm eff} \over 600}  \right)^{8/3} \left(  {1000 \over g}  \right)  {\rm ~~cm}^2/{\rm s}.
   \end{equation}
These large values of $K_{zz}$ might be regarded with some skepticism. 
 If applied to Earth's troposphere, Eq \ref{Gierasch} predicts $K_{zz}>10^7$ cm$^2$/s,
  which is two orders of magnitude too high.  Even in the present context, 
  $F_{\rm conv}$ can be much smaller than $\sigma T_{\rm eff}^4$ (or even fall to zero in places), 
  because a considerable fraction of energy transport is by radiation. 
 Depending on how well one tolerates theory, one might 
  expect $K_{zz}$ to be of the order of $10^{9}$-$10^{11}$ cm/s$^2$ in the convecting zones of self-luminous EGPs,
  and $100\times$ smaller for BDs with $g\!=\!10^5$ cm/s$^2$.
 
  It is likely that $K_{zz}$ would be much smaller in stratified gas above the convecting regions.
  On Earth, $K_{zz}$ drops by two orders of magnitude at the tropopause.
  \citet{Hubeny2007} reduce $K_{zz}$ by several orders of magnitude in the stratosphere.
  Simulations by \citet{Freytag2010} support this expectation.
  The tradeoff is between simplicity (constant $K_{zz}$) and realism (a vertical discontinuity in $K_{zz}$).
  The latter choice would introduce two more free parameters,
   the value of $K_{zz}$ in the stratosphere (ill constrained)
  and the altitude of the discontinuity (reasonably well constrained, but results could be sensitive to this). 
  A two layer model also makes interpreting the numerical results in terms
  of a quench approximation less straightforward because there can be more than one quench point.
  For this study we choose to treat $K_{zz}$ as constant with height,
  but vary it over a range wide enough to encompass all likely values.

\medskip
The chemical system used here comprises 366 forward chemical reactions and 32 photolysis reactions of 64 chemical species made of H, C, N, O, and S.  The most important missing species is probably methylamine, CH$_3$NH$_2$.
Reaction rates when known are selected from the publicly available NIST database (http://kinetics. nist.gov/kinetics).
Although many of the important reactions have been measured in both directions in the lab (e.g., 
both ${\rm CH}_4 + {\rm H} \rightarrow {\rm CH}_3 + {\rm H}_2$ and ${\rm CH}_3 + {\rm H}_2 \rightarrow {\rm CH}_4 + {\rm H}$ 
have been heavily studied),
in general one direction is much better characterized than the other.
Thermodynamic data (enthalpies and entropies) are usually better known over a wider range of temperatures.  
Thus it is better to complement each specific reaction with its exact reverse,
with the forward and reverse rates linked self-consistently by the thermodynamic data of the species involved.  
This way the transition from equilibrium to kinetically-controlled abundances is automatic.
Equilibrium is reached when all the important forward and reverse reactions are fast compared to changes in the state variables.

How this is done is fully explained by \citet{Visscher2011}. What follows is a telegraphic summary.
For reactions of the form ${\rm A} + {\rm B} \rightarrow {\rm C} + {\rm D}$ with forward reaction rate $k_f$,
the reverse rate (i.e., the rate for $ {\rm C} + {\rm D} \rightarrow {\rm A} + {\rm B}$) is $k_r=k_f\exp{\left(-\Delta G/RT\right)}$,
where $\Delta G$, the Gibbs free energy, is obtained from enthalpies and entropies of the
of the reactants and products, $\Delta G=H_A+H_B-H_C-H_D - T\left(S_A+S_B-S_C-S_D\right)$.
For associative reactions of the form ${\rm A} + {\rm B} \rightarrow {\rm AB}$, the rate for the reverse reaction (dissociation of AB) is $k_r=k_f\left(kT/P_{\circ}\right) \exp{\left(-\Delta G/RT\right)}$, where $P_{\circ}=10^6$ dynes/cm$^2$ is one atmosphere.
Similarly, the associative reverse of a dissociative reaction is given by $k_r=k_f\left(P_{\circ}/kT\right) \exp{\left(-\Delta G/RT\right)}$.

Thermodynamic data as a function of temperature 
are available for atoms and most small molecules from NIST in the form of empirical Shomate equation fits for enthalpy and entropy
(http://webbook.nist.gov/chemistry/form-ser).  
Zero point data for HS are corrected by \citet{Lodders2004}.
For CH$_3$OH and C$_2$H$_6$ we use heat capacities as a function of temperature
to derive Shomate equation fits for enthalpy and entropy.
Unfortunately, many of the more exotic free radicals are not listed in the publicly available NIST databases.
For many of these we use estimates given in \citet{Burcat2005}'s widely available gray literature compilation.  
For NNH we follow \citet{Haworth2003}, and
for N$_2$H$_2$ and N$_2$H$_3$ we follow \citet{Matus2006}.
Inaccurate thermodynamic data for free radicals are not a problem
for equilibrium calculations of well-characterized abundant species because poorly-characterized free radicals are never abundant.
On the other hand, poorly-characterized species do pose a problem in disequilibrium kinetics
because net reaction rates depend on the uncertain abundances of free radicals, 
which are determined by their thermodynamic properties. 

The lower boundary is set deep enough that all species are in thermodynamic equilibrium.
We find that pressures above 300 bars and temperatures above 2000 K usually suffice, with the limiting species
being N$_2$.
The upper boundary condition is zero flux for all species.
We place the upper boundary at $\sim\!10^{-6}$ bars, as higher altitudes require additional physics and chemistry (ion chemistry, thermospheric heating) that go beyond the scope of this work.
Steady state solutions are found by integrating the time-dependent chemical and transport equations through time using an overcorrected fully implicit backward-difference method.  
Most models take a few minutes to run to steady state from arbitrary initial conditions on a vintage laptop computer, although some particular cases can be more challenging.

Photolysis significantly affects the composition of atmospheres at very high altitudes even when the incident UV flux is small.
In our models we have set the incident UV flux to 0.1\% that at Earth.   
Photolysis plays almost no part at the higher pressures germane to quenching and is not further discussed here.
We have not included the reverses of photolysis reactions (radiative attachment, e.g.\ ${\rm OH}+{\rm H} \rightarrow {\rm H}_2{\rm O} + h\nu$)
in detailed balancing. 
Recent work has shown that radiative attachment can be important in hydrocarbon growth
 in planetary atmospheres when the resulting molecule is complex enough that radiative relaxation can be effective \citep{Vuitton2012}.

\begin{figure}[!htb] 
 \centering
 \includegraphics[width=0.45\textwidth]{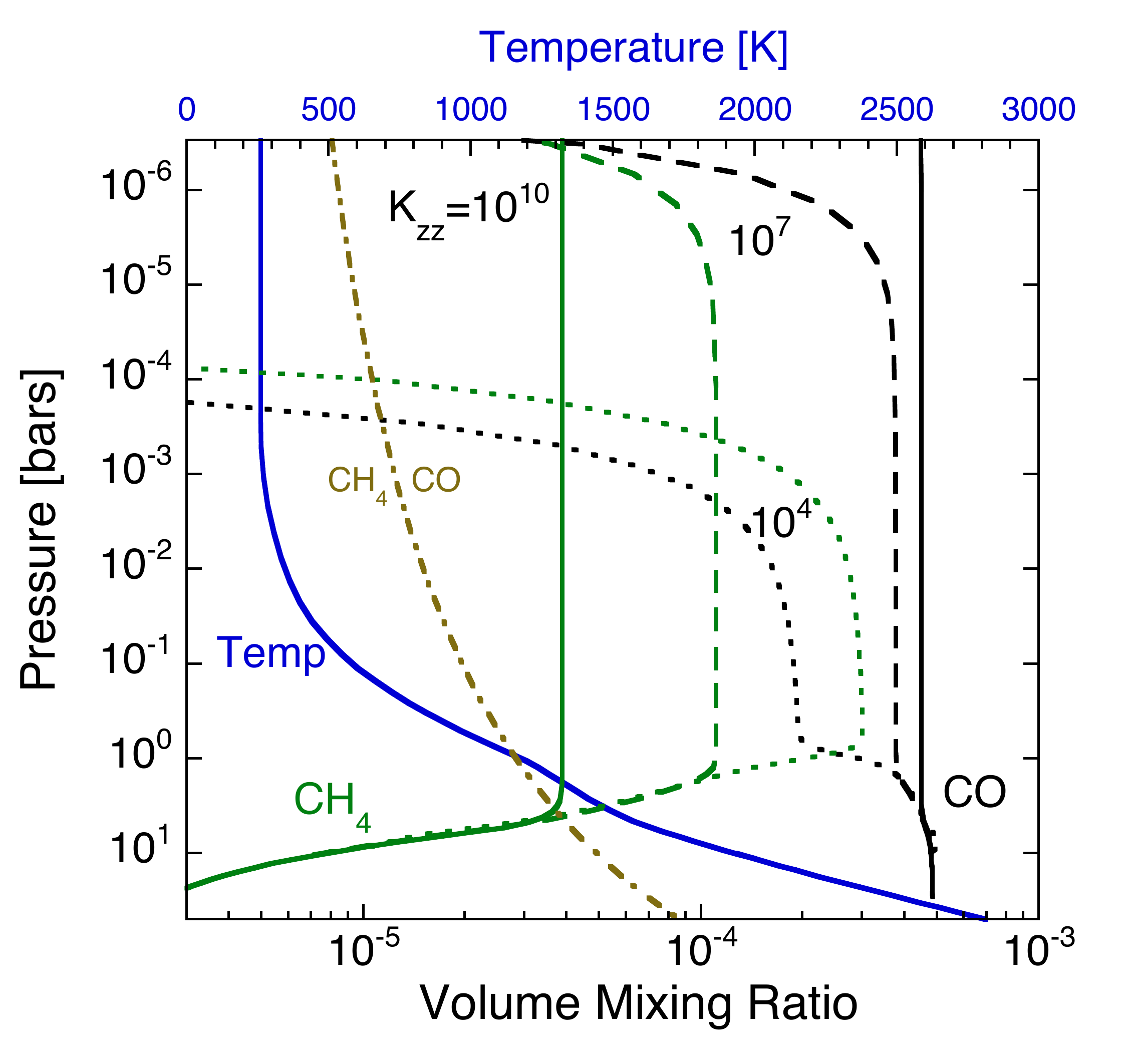} 
 
 \caption{\small Methane and CO in an exemplary lukewarm EGP.
   The planet is cloud-free, has an effective radiating temperature of 600 K, $g=10^3$ cm/s$^2$, and receives insignificant insolation. 
  The $p$-$T$ profile of the atmosphere is compared to the $p$-$T$ curve (gold dot-dash) corresponding to 
   equilibrium $f_{{\rm CH}_4}=f_{\rm CO}$.
   Methane is thermodynamically favored when $T$ is to the left of the $f_{{\rm CH}_4}=f_{\rm CO}$ curve, CO is favored to the right.
   Computed CO and CH$_4$ mixing ratios are shown from three 1D kinetics models that differ only in vertical eddy mixing:
   $K_{zz}=10^{10}$ cm$^2$/s (solid), $K_{zz}=10^{7}$ cm$^2$/s (dashed), and $K_{zz}=10^{4}$ cm$^2$/s (dots).
   Deep in the atmospheres CO and CH$_4$ are in equilibrium.  High in the atmospheres CO and CH$_4$ are unreactive and
   $f_{\rm CO}$ and $f_{{\rm CH}_4}$ would be constant but for molecular diffusion. 
   The quench points are obvious and well-defined in all three models.
  }
\label{vs_K}
\end{figure}

\subsection{Two examples}

Figure \ref{vs_K} shows how $K_{zz}$ effects chemistry on a tepid ($T_{\rm eff}=600$ K), cloud-free, relatively low mass ($g=10^3$ cm/s$^2$) extrasolar planet of solar composition. 
   Insolation is insignificant. 
   Three different $K_{zz}$ are compared.
   (i) Strong vertical mixing---$K_{zz}=10^{10}$ cm$^2$/s---suppresses CH$_4$ and maintains $f_{\rm CO} \gg f_{{\rm CH}_4}$ to very high altitudes.
    CO and CH$_4$ are in chemical equilibrium below the quench point at $\sim\! 4$ bars and 1450 K.
   (ii) A more Jupiter-like $K_{zz}=10^{7}$ cm$^2$/s raises the quench point to 1.5 bars and 1280 K, which is better for CH$_4$. 
   (iii) Weak mixing, $K_{zz}=10^{4}$ cm$^2$/s, raises the quench point to 0.6 bars and 1000 K,
   which is cool enough, barely, to fall into the CH$_4$ field,
   so that $f_{{\rm CH}_4}>f_{\rm CO}$ in this planet's photosphere.
   The effect of molecular diffusion (in which heavy molecules sink through H$_2$) 
   is apparent above $3\times 10^{-6}$ bars at $K_{zz}=10^{7}$ cm$^2$/s
   and above $3\times 10^{-3}$ bars at $K_{zz}=10^{4}$ cm$^2$/s.
   
\begin{figure}[!htb] 
   \centering
\includegraphics[width=0.45\textwidth]{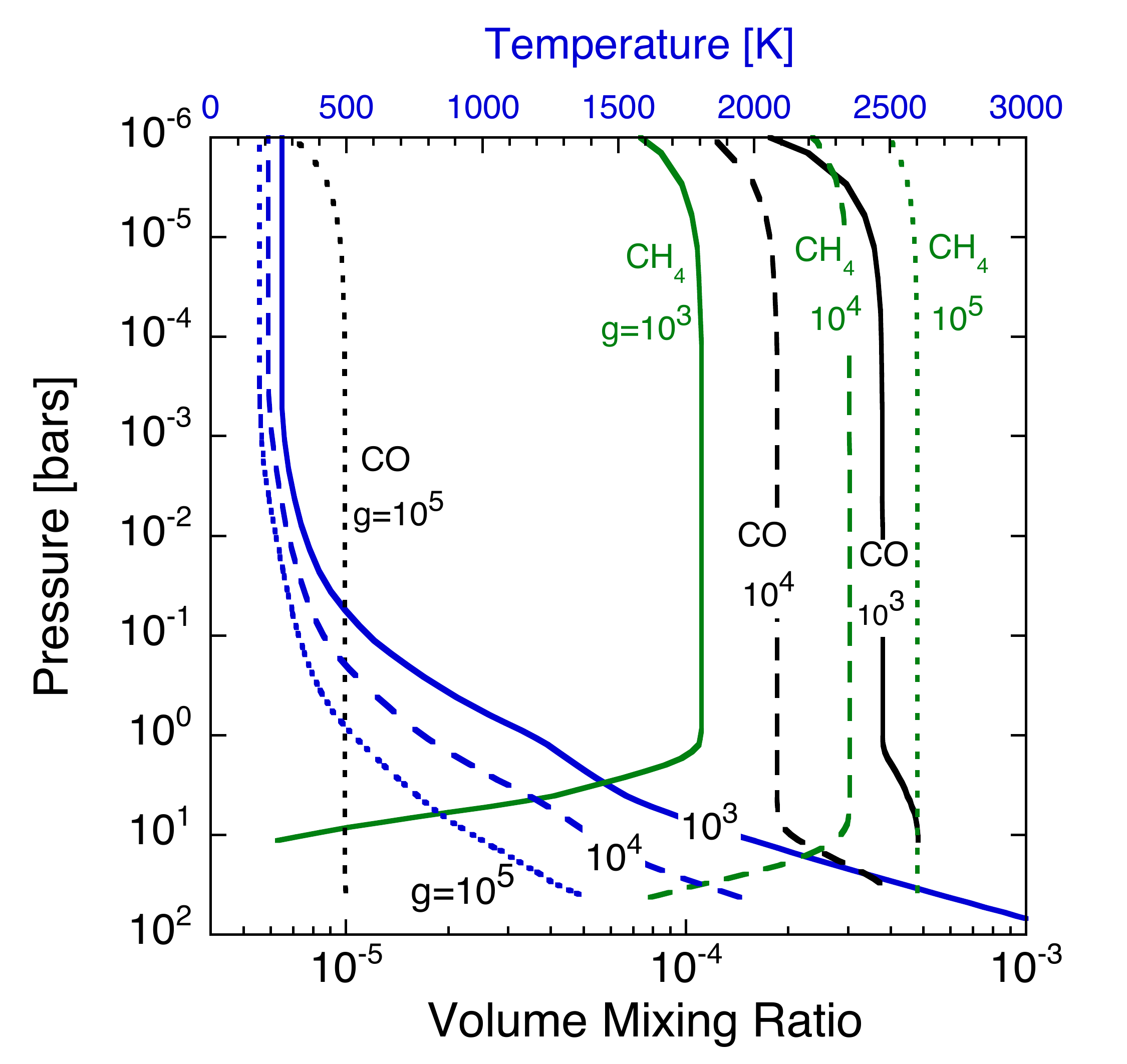} 
   \caption{\small Three models with three different surface gravities.
  The plot shows $p$-$T$ profiles and CO and CH$_4$ abundances.
  The planets are 600 K, self-luminous, cloudless, of solar composition, with 
  $K_{zz}=10^7$ cm$^2$/s. 
  Surface gravities are $g=10^3$ cm/s$^2$ (solid), $g=10^4$ cm/s$^2$ (dashed), and $g=10^5$ cm/s$^2$ (dots).
  Higher gravity models are colder at a given pressure and are more favorable to CH$_4$.
}
\label{vs_gravity}
\end{figure}

At a given effective temperature, surface gravity determines whether the atmosphere
is CO or CH$_4$ dominated.  
Figure \ref{vs_gravity} shows CO and CH$_4$ abundances in three worlds that differ only in their surface gravities.
   The atmospheres are cloud-free and of elementally solar composition.
   The models shown here have effective radiating temperatures of 600 K and a Jupiter-like $K_{zz}=10^7$ cm$^2$/s.
   Surface gravities range from a Saturn-like $g=10^3$ cm/s$^2$ to a brown-dwarf-like $g=10^5$ cm/s$^2$.
   The computed CO/CH$_4$ ratio is sensitive to surface gravity.
   This is because the $p$-$T$ profile is displaced to higher pressures and lower temperatures when gravity is higher. 
   Quenching at higher pressures and lower temperatures favors CH$_4$ over CO.
   The result is that the 600 K planet ($g=10^3$ cm/s$^2$) has abundant CO and relatively little CH$_4$,
   while the 600 K brown dwarf is methane-rich with only a trace of CO.

\subsection{The ensemble of models}

\begin{figure}[!htb] 
   \centering
\includegraphics[width=0.45\textwidth]{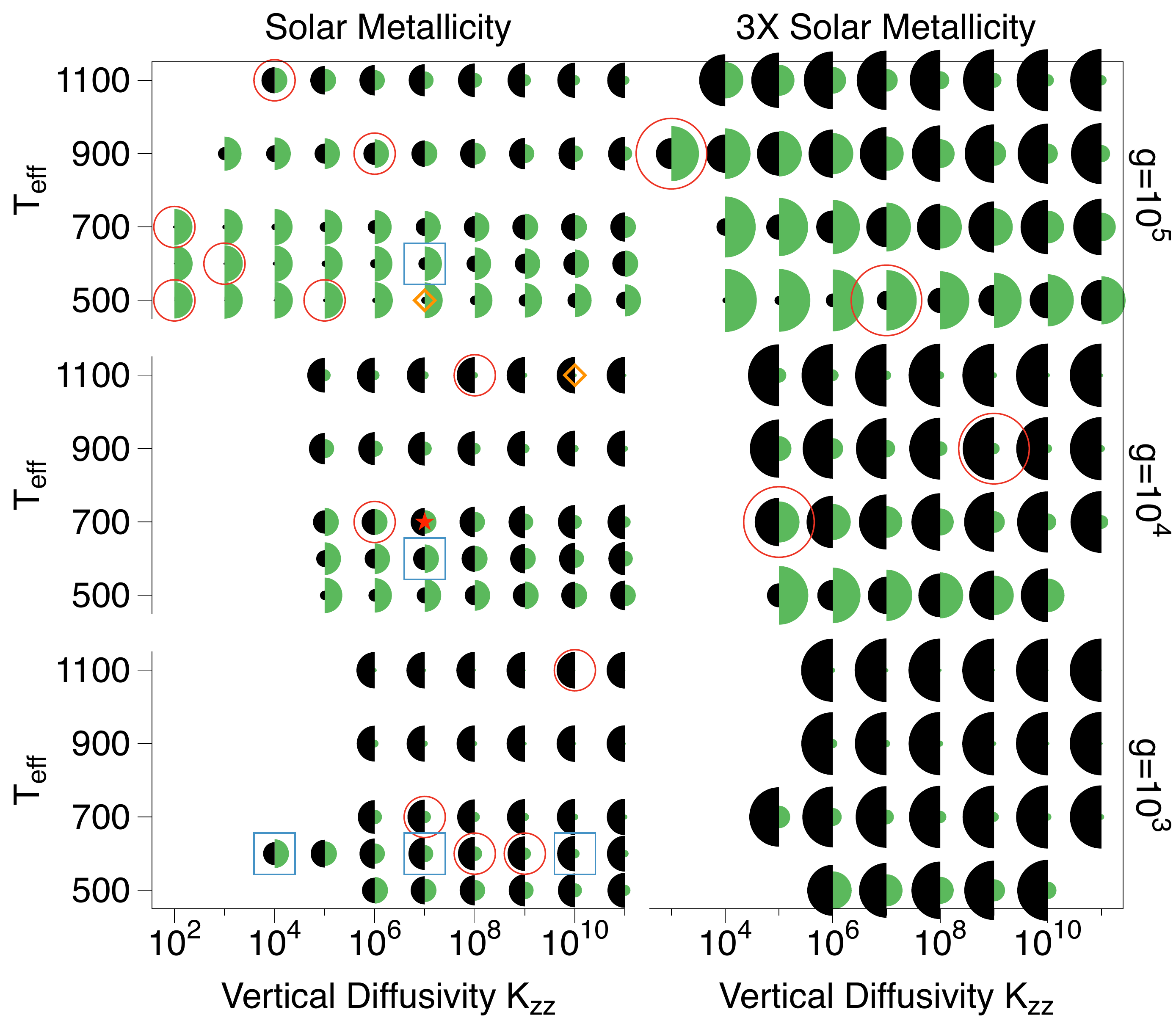} 
   \caption{\small An overview of the ensemble of models.
   There are six panels for six different combinations of gravity and metallicity. 
   Relative CO and CH$_4$ abundances above the quench point 
   are indicated by the areas of the semicircles, CO black and following, CH$_4$ green and preceding.
   In general, higher gravities, cooler temperatures, and lower metallicity favor CH$_4$ vs.\ CO.
   Different metallicities can give the same CO/CH$_4$ ratio at the same $T_{\rm eff}$ and $g$
   with modest and probably unmeasurable differences in $K_{zz}$.  
   Blue squares denote the specific cases in Figures \ref{vs_K} and \ref{vs_gravity}. 
   Red circles denote the 16 cases used in Fig \ref{YY} to illustrate quenching.
   Orange diamonds denote the two models used in Fig \ref{FF} to compare reaction rates.
   The red star is the guide to Fig \ref{chem_figure}.
   In all cases, equilibrium $f_{{\rm CH}_4} > f_{\rm CO}$.
}
\label{ensemble}
\end{figure}

We use the 1D chemical kinetics models to
explore the phase space defined by $500 < T_{\rm eff} < 1100$ K, $10^4 < K_{zz} < 10^{11}$ cm$^2$/s, and $10^3<g<10^5$ cm/s$^2$. 
Models are run to steady state.  
Ensembles were computed at solar ($m\!=\!1$) and thrice solar ($m\!=\!3$) metallicities, but
$p$-$T$ profiles were not adjusted to take into account the altered chemical composition.  
The CO and CH$_4$ abundances of all the models are summarized in Figure \ref{ensemble}. 
Low $K_{zz}$ models can be numerically challenging when the gas is very hot. 
We include two models that did not reach a true steady-state on Figure \ref{ensemble}: $T_{\rm eff}=600$ K, $g=10^3$ cm/s$^2$, $m\!=\!1$, and $K_{zz} =10^{4}$ and $10^{5}$ cm$^2$/s.
These models reached persistent near steady states, 
but were held back by undiagnosed features in the sulfur chemistry.
They are included here as part of our exploration of conditions that can produce CH$_4$ at low gravity.

\section{Determining the quench point for CO and CH$_4$}

\begin{figure*}[!htb] 
   \centering
\includegraphics[width=0.9\textwidth]{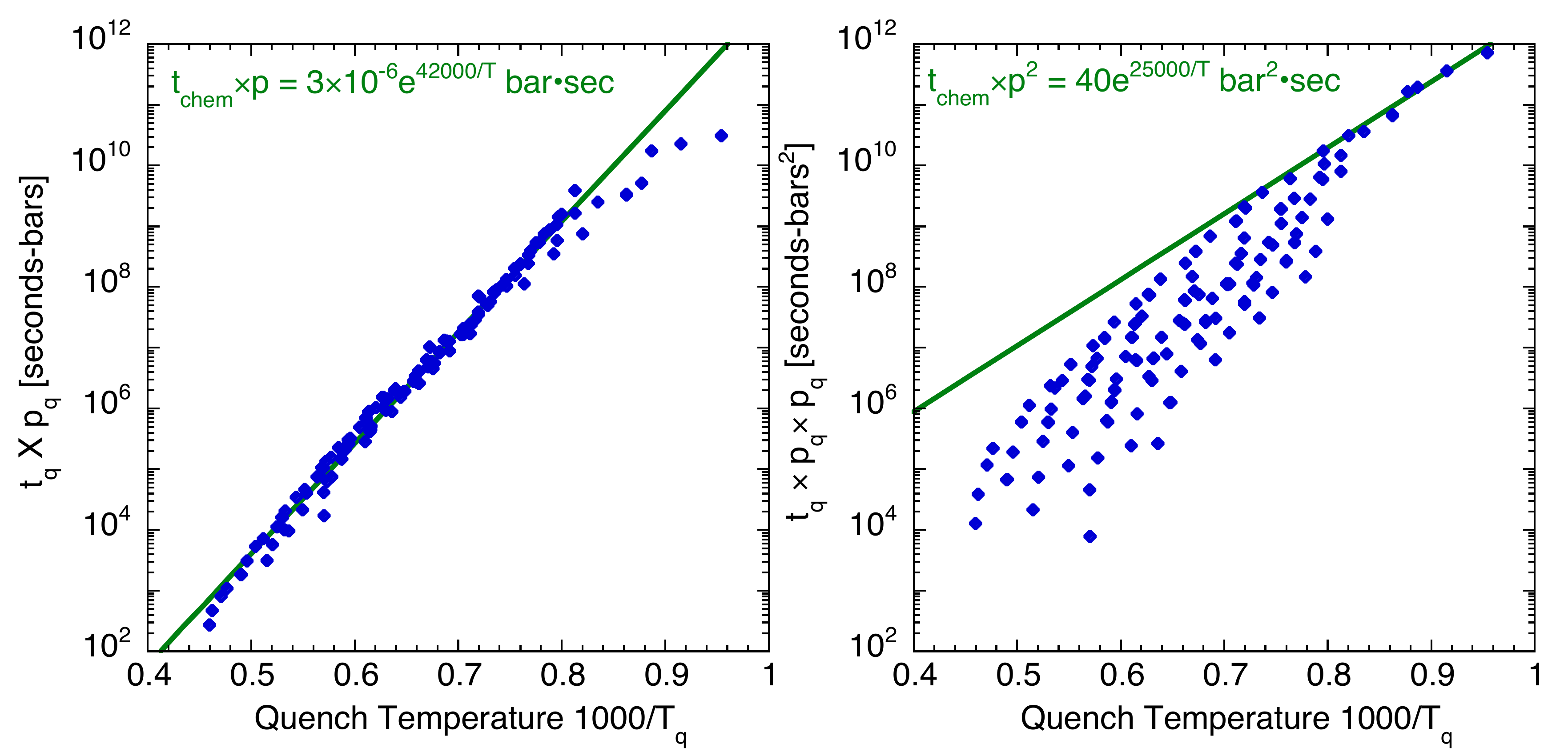}   
   \caption{\small CO-hydrogenation timescales as a function of quench point temperature $T_q$ and quench pressure $p_q$ for the ensemble of solar composition kinetics models.
  Quench conditions $T_q$ and $p_q$ are determined numerically from the full kinetics models
  using the entire network of chemical reactions.
  How this is done is explained in the text.
  The quench timescale used here is a mixing timescale,
   $t_q= t_{\rm mix}\equiv H_q^2/K_{zz}$, where $H_q$ is the pressure scale height
  at the quench point.
   The relation between the true chemical timescale and $t_q$ is arbitrary to a factor $L^2/H_q^2$, where $L$ refers to the true mixing length. 
   We do not plot $t_q$ itself (for which the points are scattered), but rather $t_q \times p_q$ ({\it left}) and
   $t_q \times p^2_q$ ({\it right}), which resolve into orderly arrays.
  {\it Left}.  A simple exponential provides a good fit to $t_q \times p_q$ for all but the coldest $T_q$. 
  {\it Right}. A different simple exponential can be fit to $t_q \times p^2_q$ for the coldest $T_q$. Lines show fits.
  }
\label{XX}
\end{figure*}

Here we want to determine the quench conditions that correspond to the asymptotic CO and CH$_4$ abundances 
in the complete kinetics models shown
in Figure \ref{ensemble}.
Our purpose in doing so is to quantify the emergent quench behaviour of the kinetics models,
and if possible describe this behaviour with a simple equation or set of equations.
That it is possible to do this is the subject of Figure \ref{XX}.
  
First, we determine the CO-CH$_4$ quench point in each atmosphere in Figure \ref{ensemble}
by finding the lowest altitude in each atmosphere where the CO and CH$_4$ mixing ratios depart measurably from equilibrium.
Deep in the atmosphere CH$_4$ and CO are in chemical equilibrium with H$_2$ and H$_2$O.
Thus the product
\begin{equation}
\label{K_test}
K_{\rm test} = { p{\rm CH}_4 \!\cdot\! p{\rm H}_2{\rm O} \!\over\! p{\rm CO} \!\cdot \! \left(p{\rm H}_2 \right)^3 }  
\end{equation}
equals the corresponding equilibrium constant 
\begin{equation}
\label{K_eq}
K_{{\rm CH}_4\cdot {\rm CO}} = 5.24 \times 10^{-14} \exp{\left(27285/T\right)}.
\end{equation}
In Eqs \ref{K_test} and \ref{K_eq} (and all succeeding expressions of similar form), pressures are in atmospheres and temperatures in Kelvins.

In the regions of a convective atmosphere relevant to the present investigation,
equilibrium contours of constant CH$_4$/CO
are always shallower than the adiabat, hence the equilibrium CH$_4$ mixing ratio always decreases with increasing depth.
To show this (and its limits), 
 use the adiabatic relation $T^{\,\gamma} \propto  p^{\,\gamma -1}$,
where 
$\gamma = C_p/C_v $ has its usual meaning as the ratio
of heat capacities.
The temperature gradient along the adiabat (constant entropy $S$) is 
\begin{equation}
\label{adiabatic_gradient}
\left({\partial T\over \partial p}\right)_{\! S} =  {\gamma -1 \over \gamma} {T\over p} = {R \over C_p} {T\over p} .
\end{equation}
For a temperate cosmic mix of H$_2$ (84\%) and He (16\%), we can expect H$_2$ to have $C_p\approx 3.5 R$ and He to have $C_p=2.5 R$, and hence $R/C_p \approx 0.30$. 
Warmer temperatures excite hydrogen's vibrational modes, for which $C_p\approx 4.5 R$ and $R/C_p \approx 0.24$.
Detailed calculations show that $R/C_p$ decreases
monotonically from $0.29$ at 1000 K to $0.25$ at 2100 K to $0.20$ at 3000 K as dissociation (another energy sink) becomes important.

Equation \ref{adiabatic_gradient} needs to be compared to contours of equilibrium chemistry.
Defining $\xi \equiv {\rm CH}_4/{\rm CO}$,  
we can write the equilibrium (Eqs \ref{K_test} and \ref{K_eq}) in the form
\begin{equation}
\label{K_eq2}
\xi \!\cdot \!\left( f_{\rm O} - { f_{\rm C} \over 1+\xi} \right) = p^2 f_{{\rm H}_2}^3 A e^{B/T}
\end{equation}
where $f_{{\rm H}_2}\approx 0.84$ is the H$_2$ mixing ratio, $f_{\rm O} \approx f_{{\rm H}_2{\rm O}} + f_{\rm CO}$ is the total oxygen mixing ratio, and $f_{\rm C} \approx f_{{\rm CH}_4} + f_{\rm CO}$
is the total carbon mixing ratio. 
These are generally very good approximations for $T<2500$ K.
The quantity in parentheses is $f_{{\rm H}_2{\rm O}}$.   
For contours of constant $\xi$, the left hand side of Eq \ref{K_eq2} is constant.
Taking derivatives,
the temperature gradient is 
\begin{equation}
\label{K_eq3}
\left({\partial T\over \partial p}\right)_{\!\xi} = {2T\over B} \, {T \over p}  \approx 0.11 {T \over 1500} \, {T \over p} .
\end{equation}
Thus $\left({\partial T / \partial p}\right)_{\xi} < \left({\partial T / \partial p}\right)_S$ for temperatures lower than about 2900 K,
by which point several of the assumptions made in deriving this relation have begun to break down.
Even with $K_{zz}$ as high as $10^{11}$ cm$^2$/s, quenching in the CO-CH$_4$ system takes place at temperatures less than 2200 K.
Hence we can always presume that CO is most abundant and CH$_4$ least abundant at the lower boundary of our model.
It follows that we can determine the quench conditions (temperature $T_q$, pressure $p_q$)
by working up from the lower boundary and finding the lowest altitude at which the 
product $K_{\rm test}$ is appreciably smaller than the equilibrium constant $K_{{\rm CH}_4\cdot {\rm CO}}$.
In practice we searched for the lowest altitude where $K_{\rm test}<0.9 K_{{\rm CH}_4\cdot {\rm CO}}$.

\medskip
We also wish to determine a quench timescale $t_q$. 
In principle, the right way to do this is to determine the true chemical reaction timescale from the sum of
all relevant reaction rates at the quench point. 
In general this information is surprisingly hard to extract from a model, 
because the breakdown of equilibrium appears as small differences 
between much larger forward and reverse reaction rates in many reactions.   
This approach is practical only if one already knows exactly which species and reactions are key.
Once known, for $t_q$ to be useful in a quench approximation one then needs a comparably fine-grained 
description of the cooling time implicit in $K_{zz}$ and the particular chemical reactions that establish $t_q$.

Much easier, and more useful, is to define a quench timescale for the system as a whole 
in terms of a relevant mixing time,
 $t_q= t_{\rm mix} = L^2/K_{zz}$
 where $L$ is a characteristic length scale.
The simple choice is to set $L=H_q$, the scale height at $T_q$.
This definition is arbitrary to within a multiplicative factor  
because the true mixing length $L$ is not in general equal to $H_q$.  
But this definition of $t_q$ is precisely what is needed
for constructing a new quench approximation that assumes $t_{\rm mix}= H_q^2/K_{zz}$ and,
as seen in Figure \ref{XX}, it leads to a very simple description of
 the overall behaviour of the full network of chemical reactions. 
 
\citet{Smith1998} demonstrated that $L$ is often considerably shorter than $H$,
mostly because the key reactions determining $t_q$ are much more sensitive to temperature than to pressure. 
\citet{Smith1998} developed a five step iterative algorithm for determining $L$ and presented it in the form of a ``recipe'' to be used
in quenching calculations,
which is especially useful when the system is controlled by a single reaction whose rate is known.
Several workers have employed Smith's recipe.
Unfortunately, the range of mixing times that results can be very great.
For example, \citet{Visscher2011} 
list values for $L$ between $0.1\,H$ and $0.7\,H$ for the CO-CH$_4$ system, 
the equivalent of a factor of 50 in $t_{\rm mix}$
and therefore equivalent to a factor of 50 in the inferred value of $K_{zz}$, itself a parameter of easy virtue.
The resulting uncertainty frustrates intercomparison between models. 
Our position is that the most useful approximations are simple approximations.
We therefore stick with the old rule $L=H$ to define $t_{\rm mix}$, rather than iterate between $L$ and $T_q$.
How $L$ is defined is important for comparing to previous work that tries
to estimate $t_{\rm chem}$ from particular reaction rates, but it is simply a scaling factor for us. 
 
Irrespective of how $L$ is defined,
we find that we can fit the quench time $t_q$ to a simple functional form that spans all our models.
We start from the expectation that ${\rm CO}\rightarrow {\rm CH}_4$ will depend on temperature and pressure, perhaps the C/O ratio, and possibly on metallicity.
It is reasonable to hope that the net reaction has an Arrhenius-like rate
\begin{equation}
\label{XV}
t_{q} = A \; p^{-b} m^{-c} \exp{\left(B/T\right)},
\end{equation}
where possible pressure and metallicity dependencies are written as power laws.

Our hope is nicely borne out by Figure \ref{XX}, which
plots $t_{q}$ for the CO-CH$_4$ reaction
for our complete roster of solar metallicity models.
All but the coldest quench points are described by a rather well-defined Arrhenius expression with $b\!=\!1$.
The Arrhenius fit (for $m\!=\!1$) is 
\begin{equation}
\label{t_q_0} 
 t'_{q1} = 3.0\times 10^{-6} p^{-1} \exp{\left(42000/T\right)} {\rm ~sec,}    
\end{equation}
where $p$ is in bars.   
The time scale for models with $m\!=\!3$ (not shown) is roughly half of this.

\begin{figure*}[!htb] 
   \centering
\includegraphics[width=0.9\textwidth]{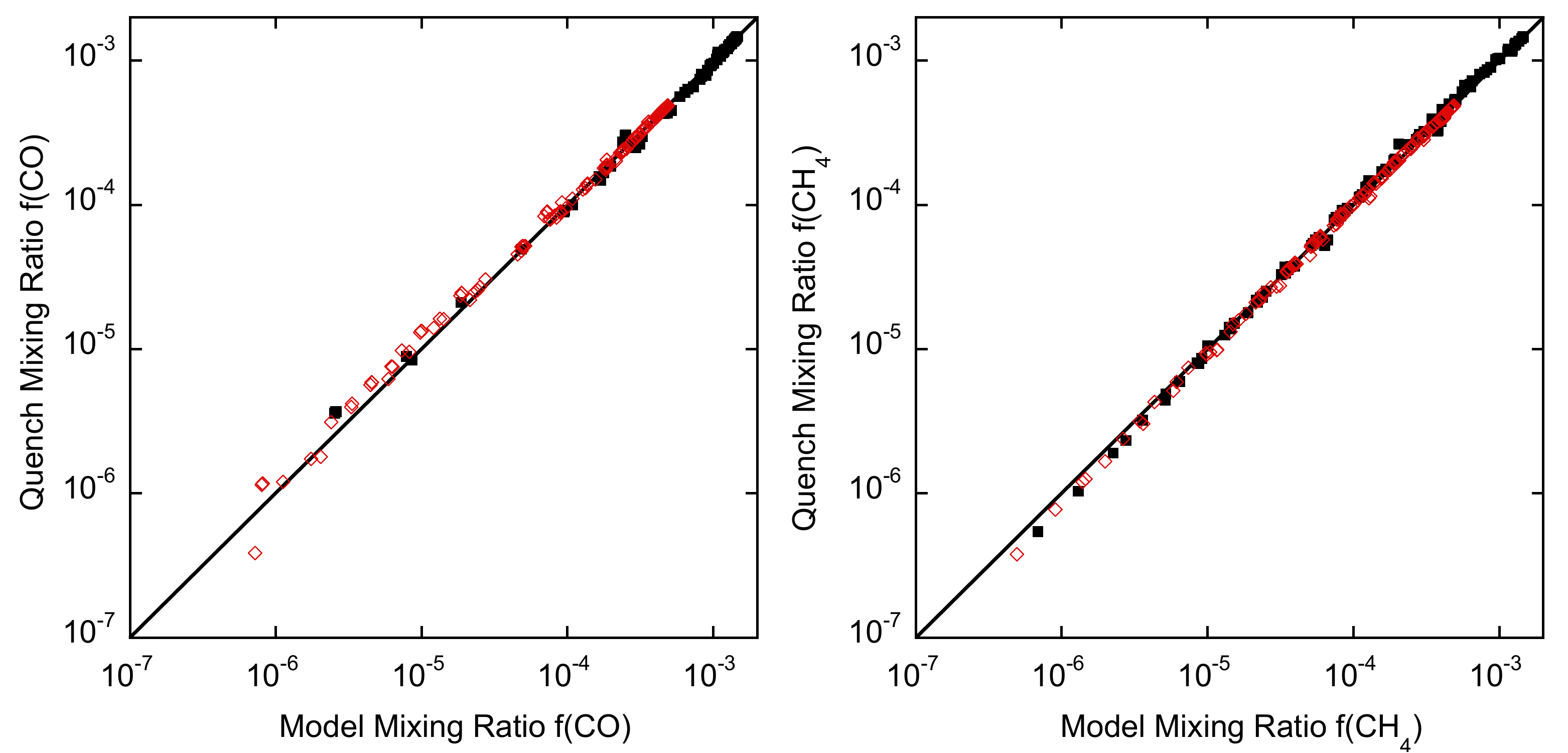}
\caption{\small CO (left) and CH$_4$ (right) mixing ratios predicted by the quenching approximation $t_{\rm chem}=t_{\rm mix}$
(y-axis)
plotted against the 
actual 
asymptotic mixing ratio determined from the full ensemble of 1-D kinetics model
(x-axis). A perfect approximation would adhere to the line.
 Open red symbols are for $m\!=\!1$ and solid black symbols are for $m\!=\!3$ using Eq \ref{t_q_1}. }
\label{VV}
\end{figure*}

The prime is placed on $t_{q1}'$ because, 
by using $K_{\rm test}<0.9 K_{eq}$ rather than $K_{\rm test}< K_{eq}$, we have 
overestimated the quench time.
 To correct for this---and to verify that a quench scheme works---we compare
the predictions of the quench scheme with the predictions of the ensemble of models.
Fits are shown in Figure \ref{VV}.  These use
\begin{equation}
\label{t_q_1} 
 t_{q1} = 1.5\times 10^{-6} \, p^{-1} m^{-0.7} \exp{\left(42000/T\right)} {\rm ~sec.}  
\end{equation}
The methods of Figure \ref{VV} and Figure \ref{XX} are complementary.
A satisfactory fit on Figure \ref{VV} is more sensitive to the $A$ factor (overall rate) and the metallicity (Eq \ref{XV}),
whilst the fit on Figure \ref{XX} is more sensitive to the $B$ factor (temperature dependence) and pressure.

The half-dozen models that are ill fit by Eq \ref{t_q_0} are fit instead by a second Arrhenius relation
\begin{equation}
\label{t_q_2} 
 t_{q2} = 40 \, p^{-2} \exp{\left(25000/T\right)} {\rm ~sec.}   
\end{equation}
where we have given the iterated rate.
The models fit by Eq \ref{t_q_2} are those with very low quench temperatures $T_q$ and very little CO.  
We have no such models with $m\!=\!3$.

The high and low temperature fits combine harmonically to give a general description of reaction times
in the CO-CH$_4$ system,  
\begin{equation}
\label{t_chem}
t_{\rm CO} = \left( {1\over t_{q1}} + {1\over t_{q2}} \right)^{-1}.
\end{equation}
Equation \ref{t_chem} when used with the timescale $t_{\rm mix}  = H^2/K_{zz}$ 
works for all of our models
for CO and CH$_4$ and we expect should work generally for compositions not too far from solar.

\subsection{Analysis}

\begin{figure*}[!htb] 
   \centering
\includegraphics[width=0.8\textwidth]{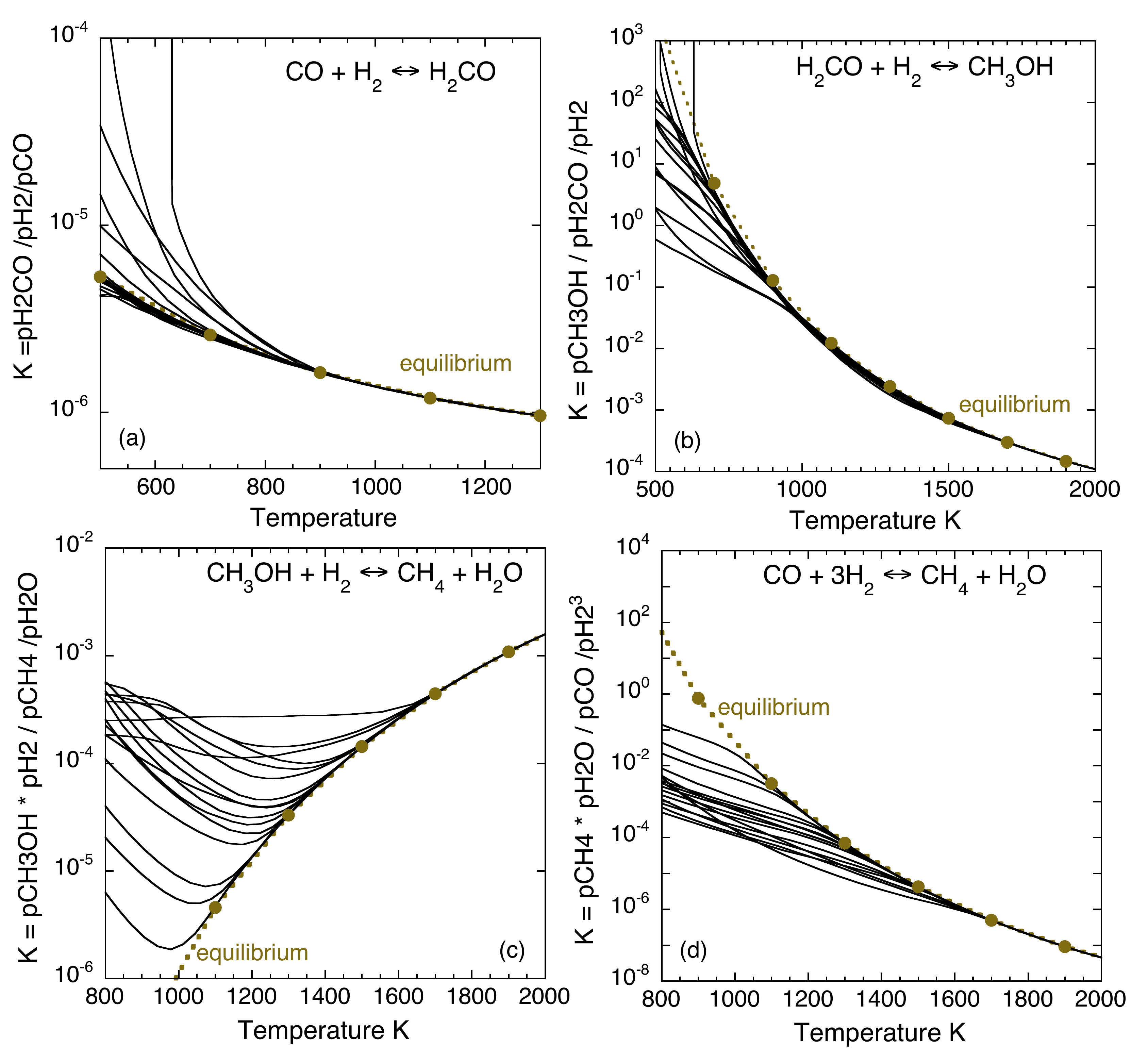} 
   \caption{\small Quenching behavior of the three intermediate reactions (a, b, c) and of the ${\rm CO} \leftrightarrow {\rm CH}_4$ system as a whole (d).
  Results are shown from 16 models chosen to provide a representative sample of the ensemble (marked by red circles in Fig. \ref{ensemble}).
     Panels (a,b) show that ${\rm CO} \leftrightarrow {\rm H}_2{\rm CO}$ and ${\rm H}_2{\rm CO}  \leftrightarrow {\rm CH}_3{\rm OH}$ appear to be
   equilibrated in all models at $T > 900$ K.
   Panel (c) shows that equilibration of ${\rm CH}_3{\rm OH}$ and ${\rm CH}_4$ is more difficult.
   In detail the quench points for ${\rm CH}_3{\rm OH} \leftrightarrow {\rm CH}_4 $ appear to be the same as the quench points
   for ${\rm CO} \leftrightarrow {\rm CH}_4$ as shown in panel (d) for the same 16 models.
   Evidently the bottleneck is between ${\rm CH}_3{\rm OH}$ and ${\rm CH}_4 $, 
    in accord with conclusions reached by Moses et al (2011). }
\label{YY}
\end{figure*}

We examine the ensemble of models   
for quench points pertinent to the reaction intermediates, to determine where in the CO-CH$_4$ 
reaction network quenching first takes place.
We consider the CO-H$_2$CO, H$_2$CO-CH$_3$OH, and CH$_3$OH-CH$_4$ equilibria.
We analyzed 16 models taken arbitrarily from the ensemble. The chosen ones are
marked on Fig \ref{ensemble} with scarlet circles.
We find, to within our ability to determine this, that the CH$_3$OH-CH$_4$ reaction
quenches at the same temperature as the CO-CH$_4$ reaction as a whole.
This is shown by panels (c) and (d) of Figure \ref{YY}.
By contrast the CO-H$_2$CO and H$_2$CO-CH$_3$OH reactions stay near equilibrium for
temperatures above 900 K (Figures \ref{YY}a and \ref{YY}b).
Hence as a parcel cools equilibrium first breaks down between CH$_3$OH and CH$_4$.
This is in accord with conclusions reached 
by \citet{Moses2011} and \citet{Visscher2011}.

\begin{figure*}[!htb]
\plottwo{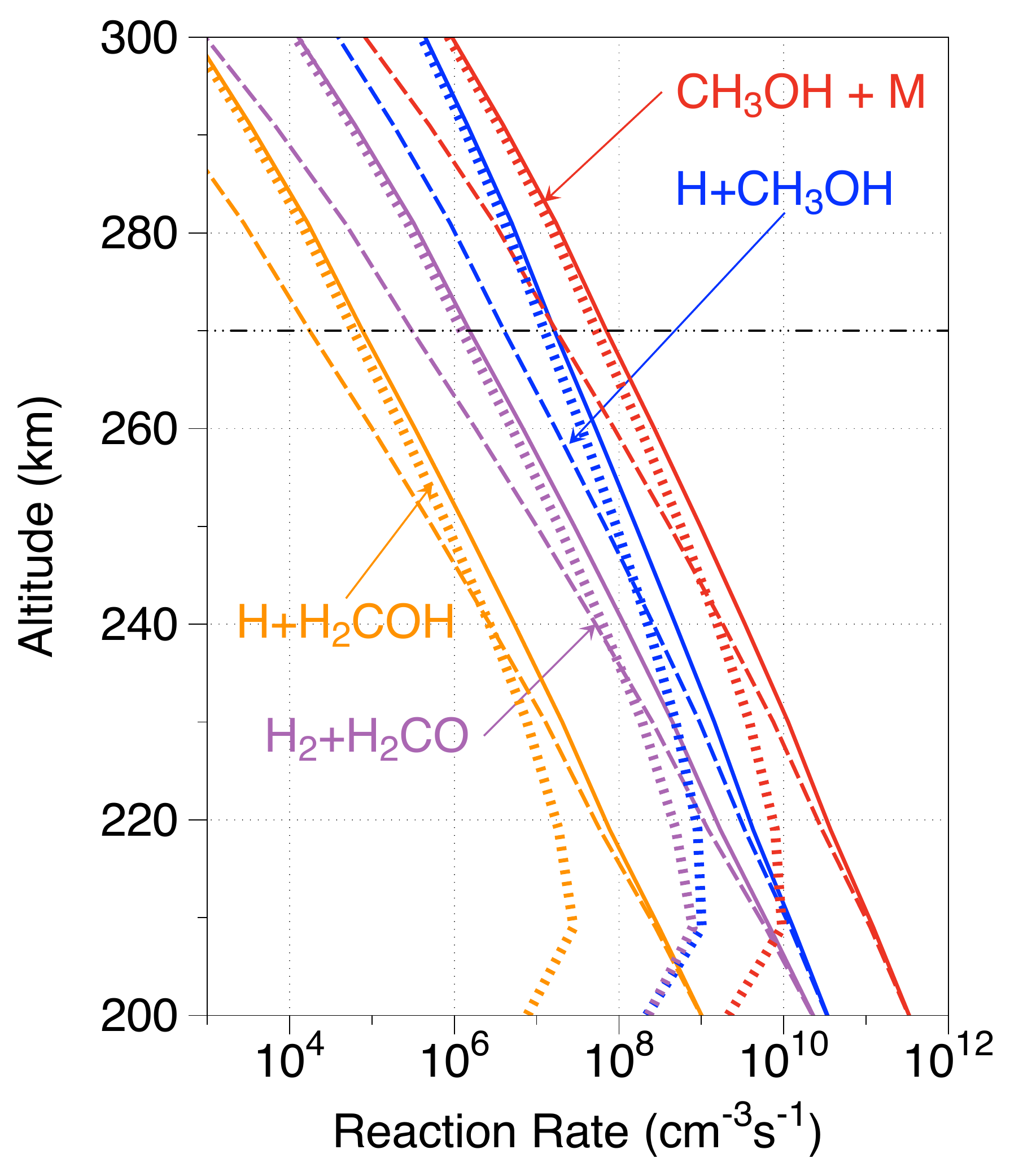}{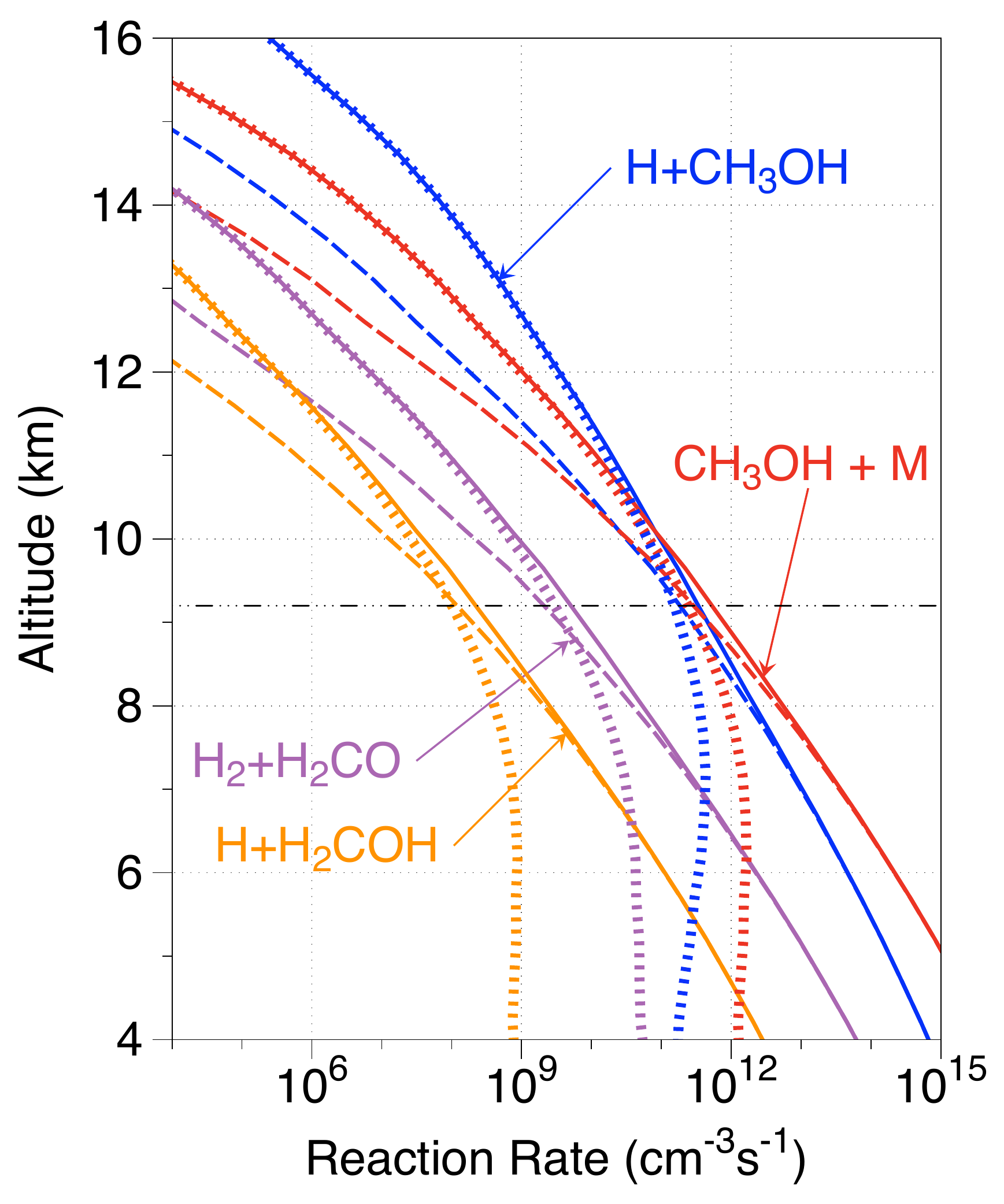} 
  \caption{\small The four most important reactions for quenching in our model CO-CH$_4$ system.
  All four reactions yield ${\rm CH}_3+{\rm OH}$ as products.
  Solid curves show rates of the forward reaction that creates CH$_3$,
  dashed curves show the reverse reaction that destroys CH$_3$,
  and wide dots the difference (the net production of CH$_3$), which is equivalent to net loss of CO.
  Black dot-dash lines are the altitudes where CH$_4$ and CO reach their final quenched values.
   {\it Left:} A warm, strongly-mixed ($T_{\rm eff}, K_{zz}, g =1100, 10^{10}, 10^4$), 
 model dominated by CO.  Disequilibrium reactions are fastest at 210 km (at 1700 K) but 
 full quenching is delayed until 270 km (at 1360 K);
 the distance corresponds to a bit more than a pressure scale height.
  {\it Right:} A cool, moderately-mixed ($T_{\rm eff}, K_{zz}, g =500, 10^{7}, 10^5$) 
 BD dominated by CH$_4$.  Disequilibrium reactions are fastest at 7 km but quenching is at 9 km
 (the scale height is 5 km).
  }
\label{FF}
\end{figure*}

Figure \ref{FF} examines in detail the key reactions that break the CO-CH$_4$ equilibrium.
The important reactions form the methyl radical CH$_3$ from
species such as CH$_3$OH in which the C-O bond is still intact. 
Figure \ref{FF} illustrates two cases, one a cool ($T_{\rm eff}=500$ K) brown dwarf with a preference for
 CH$_4$ ($f_{\rm CO}=1.4\times 10^{-5}$, $f_{{\rm CH}_4}=4.7\times 10^{-4}$),
the other a warmer ($T_{\rm eff}=1100$ K), smaller world that favors
CO ($f_{\rm CO}=4.8\times 10^{-4}$, $f_{{\rm CH}_4}=1.0\times 10^{-5}$).
  Both models assume solar metallicity ($m\!=\!1$). 
In both cases the most important reaction is simple thermal decomposition of methanol,
${\rm CH}_3{\rm OH} + {\rm M} \rightarrow {\rm CH}_3 + {\rm OH}$,
although the reaction of methanol with atomic hydrogen is nearly as fast in the colder,
higher gravity case. The reaction ${\rm H}_2{\rm COH} + {\rm H}$, which \citet{Moses2011} report as the most important in their CO-rich models, does not appear to be very important in ours.
We do not claim that our system is better;
we merely point out that abundances and reaction rates of furtive species like ${\rm H}_2{\rm COH}$ are
highly uncertain.

\subsection{Comparisons with previous work}

To put our results in perspective, we compare them to some quench approximations  
 seen in the literature.  
 Published quench approximations begin by identifying a particular forward reaction 
 as the bottleneck and then calculate CO loss time scales from the chosen reaction's rate;
 variety lies in the reactions that are chosen and the reaction rates adopted. 
 
\begin{figure*}[!htb]
\plottwo{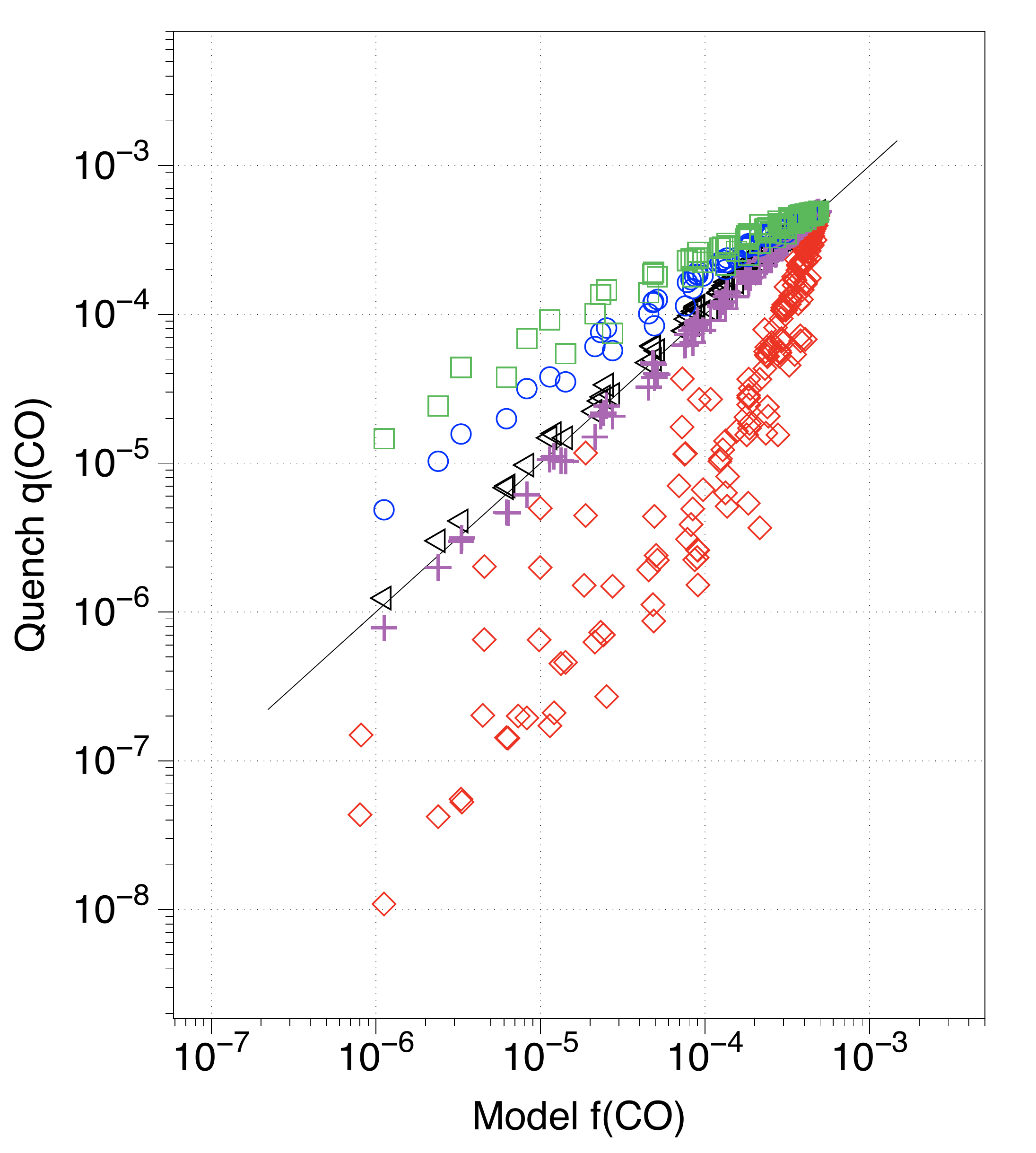}{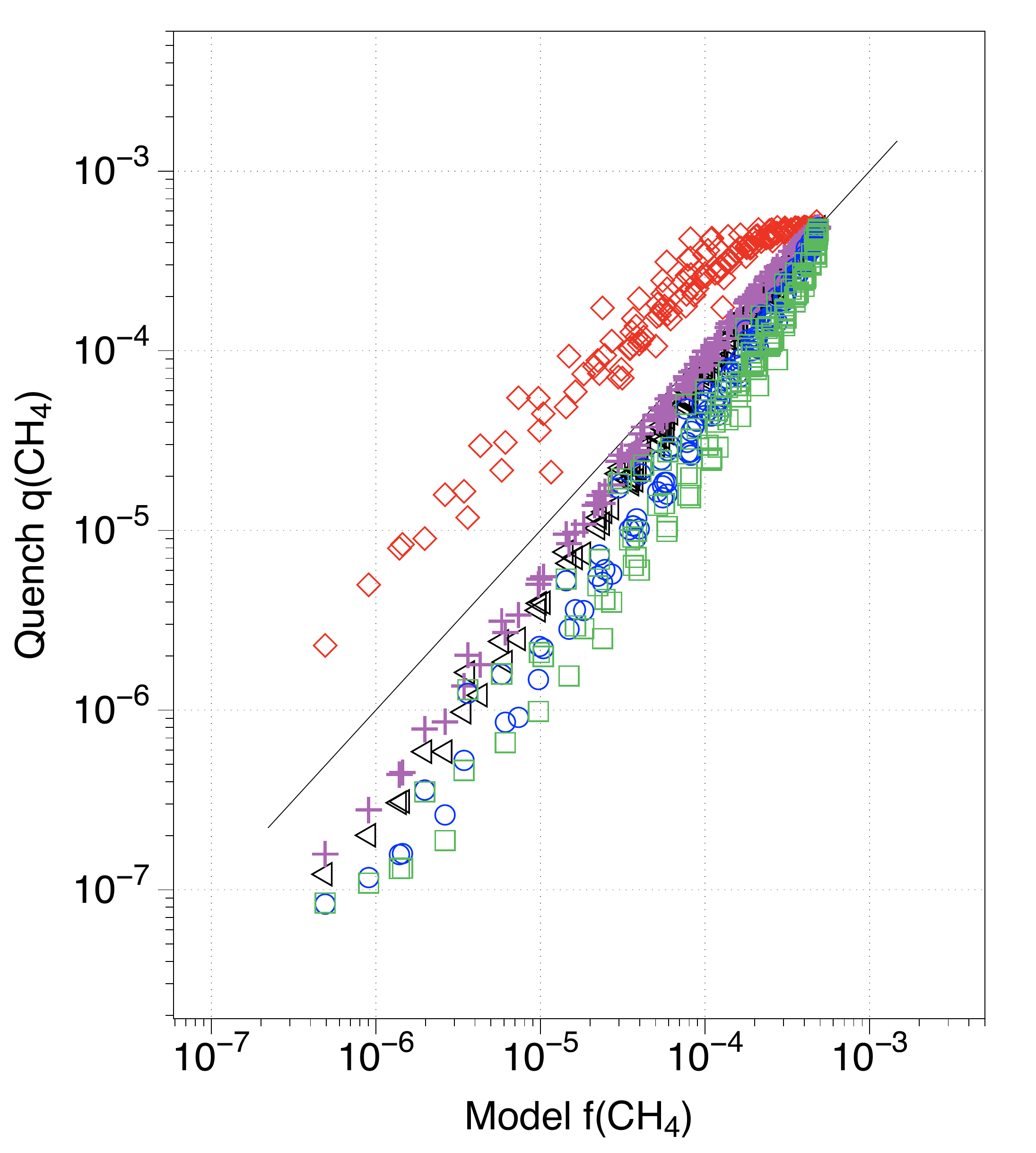}
  \caption{\small Comparison of some quench approximations 
 (y-axes)
  to the
  actual
  CO and CH$_4$ mixing ratios computed by the full 1D model
  (x-axes). A perfect approximation would adhere to the line.
  Only results for solar metallicity ($m\!=\!1$) are shown.
  Purple plusses follow the widely applied \citet{Prinn1977} model with $L=H$.
  Red diamonds use the channel through CH$_3$O suggested by \citet{Yung1988} 
  but using a more recent estimate for the reaction rate and reducing the mixing length to $L=0.14\,H$;
  this is not the same as \citet{Yung1988}'s algorithm.  
  Green squares implement \citet{Line2011}'s CH$_3$O channel with $L\!=\!H$.
  Black triangles implement the methanol decomposition channel favored for low $f_{\rm CO}$
  by \citet{Moses2011} and \citet{Visscher2011}.  These are computed with mixing length $L=0.14\,H$.  
  Blue circles represent our attempt to implement the \citet{Moses2011} H$_2$COH channel with $L=0.14\,H$.}
  \label{PB}
\end{figure*}

 We begin with the classic \citet{Prinn1977} prescription,
used by \citet{Hubeny2007} as their ``slow'' case.
\citet{Prinn1977} suggested  
\begin{equation}
\label{prinn}
  {\rm H}_2 + {\rm H}_2{\rm CO}  \rightarrow {\rm CH}_3 + {\rm OH}
\end{equation} 
as a rate-limiting reaction,
with reaction rate $k_{f}=2.3\times 10^{-10} e^{-36200/T}$ cm$^3$/s. 
R\ref{prinn} is a reaction that jumps over two energy barriers in Figure \ref{chem_figure},
which makes it look like it should be relatively unlikely.
Figure \ref{FF} shows that R\ref{prinn} is modestly important in our system,
but we use a newer slower rate estimated by Jasper et al (2007). 
The time constant for CO loss by reaction R\ref{prinn} using \citet{Prinn1977}'s rate
is approximated by the forward reaction  
 \begin{equation}
   t_{\rm chem}^{-1} = {-1\over \left[{\rm CO}\right]}  {\partial \left[{\rm CO}\right] \over \partial t} = {k_{f} \left[{\rm H}_2\right] \left[{\rm H}_2{\rm CO}\right] \over  \left[{\rm CO}\right] } {\rm ~sec}^{-1}
\end{equation} 
where the notation $\left[ {\rm CO} \right] $ refers to number density [cm$^{-3}$].  
To evaluate $t_{\rm chem}$ for R\ref{prinn}, we put 
formaldehyde in equilibrium with CO and H$_2$,
\begin{equation}
\label{formaldehyde}
  {p{\rm H}_2{\rm CO}\over p{\rm H}_2 \!\cdot\! p{\rm CO} } = K_{{\rm H}_2{\rm CO}} = 3.3\times 10^{-7} \exp{\left(1420/T\right)} ,
\end{equation}  
to obtain
  \begin{equation}
 \label{t_pb}
  t_{\rm chem} = {1.3\times 10^{16} \exp{\left(34780/T\right)} \over p\, N f_{{\rm H}_2}^{2}} {\rm ~sec}.
\end{equation} 
The number density $N$ is related to the pressure $p$ in bars by $NkT\!=\!10^6\,p$.
 The predictions of Eq \ref{t_pb} are compared to results from our complete models in Figure \ref{PB}.
  Equation \ref{t_pb} works quite well in a quench approximation to our full model, especially for cases where CH$_4$ is predicted to be abundant, despite its being based on the wrong reaction with the wrong rate. 
  As the \citet{Prinn1977} approximation has been widely used for a very long time, it is valuable to see
  that it seems to work rather well. 
  When compared to Eq \ref{t_q_1}, Eq \ref{t_pb} has a stronger pressure dependence ($p^{-2}$ vs.\ $p^{-1}$),
  a weaker temperature
 dependence ($T\!\cdot\! e^{34780/T}$ vs.\ $e^{42000/T}$), and no dependence on metallicity.  

More recent discussions of the CO-CH$_4$ quench approximation omit enough details that 
they can be challenging to reproduce.  
We attempt to do so here because the comparisons are illuminating.   
\citet{Yung1988} suggested that the rate-limiting step is the 3-body reaction 
\begin{equation}
\label{yung}
  {\rm H} + {\rm H}_2{\rm CO} + {\rm M} \rightarrow {\rm CH}_3{\rm O} + {\rm M}.
\end{equation} 
\citet{Hubeny2007} use R\ref{yung} as their ``fast'' rate with $L\!=\!H$.
\citet{Barman2011a} also use this rate, but following \citet{Smith1998} they set the mixing length $L$ to  
unspecified values between 10\% to 20\% of the scale height, which effectively slows the ``fast'' rate by a factor of 25-100
 compared to $L\!=\!H$.
\citet{Cooper2006} implemented the CH$_3$O channel with a different reaction rate
while following Smith's ``recipe.''   
We use the asymptotic high pressure rate
 
$k_{\infty} = 8.0\times 10^{-10}\exp{\left(-3160/T\right)}$,
 
a very fast rate which we obtained by reversing the first order rate for
${\rm CH}_3{\rm O} \rightarrow {\rm H} + {\rm H}_2{\rm CO}$ given by \citet{Rauk2003}. 
If we presume that H and H$_2$CO are both in equilibrium, 
we can use the hydrogen equilibrium constant $K_{{\rm H}\cdot {\rm H}_2}=1155\, e^{-26916/T} = p{\rm H}\cdot p{\rm H}_2^{-0.5}$
to obtain
  \begin{equation}
  t_{\rm chem}^{-1} = {k_{\infty} \left[{\rm H}\right] \left[{\rm H}_2{\rm CO}\right] \over  \left[{\rm CO}\right] } = 
   k_{\infty}\,  p^{1.5} \,f_{{\rm H}_2}^{1.5}  \left( 10^6\over kT\right) K_{{\rm H}_2{\rm CO}} K_{{\rm H}\cdot {\rm H}_2}. 
\end{equation} 
Evaluated,
  \begin{equation}
\label{t_y}
t_{\rm chem} = 4.5\times 10^{-10}\, T p^{-1.5} f_{{\rm H}_2}^{-1.5}   \exp{\left(28656/T\right)}.
\end{equation} 
Results of using Eq \ref{t_y} are shown as red diamonds on Fig \ref{PB} with $L=0.14H$, which is comparable
to what \citet{Barman2011a} use (the fit would look worse with $L=H$).     
These stand out with too much CH$_4$ and too little CO,
 because R\ref{yung} is not a true bottleneck.

\citet{Line2011} moved the bottleneck to ${\rm CH}_3{\rm O} + {\rm H}_2 \rightarrow {\rm CH}_3{\rm OH}+ {\rm H}$.
To compute a rate constant, 
we use $p{\rm CH}_3{\rm O} = K_{{\rm CH}_3{\rm O}} \!\cdot\! p{\rm H}_2 \!\cdot\! p{\rm H} \!\cdot\! p{\rm CO}$,
where $K_{{\rm CH}_3{\rm O}} = 3.0\times 10^{-11}e^{10582/T}$ is the relevant equilibrium constant
(the large number of significant digits in these equilibrium constants do not imply accuracy).  
To complete the equation we also assume that H and H$_2$ are in equilibrium.
We use the reaction rate given by \citet{Line2011}, which results in 
\begin{equation}
\label{line_CO}
t_{\rm chem} = {1.2\times 10^{29} \exp{\left(45720/T\right)} \over T^4 N p  f_{{\rm H}_2}^{2} } {\rm ~~sec}.
\end{equation}
Results are shown as green squares on Figure \ref{PB} using $L=H$, as \citet{Line2011} do.
Because it predicts too much CO and too little CH$_4$, $t_{\rm chem}$ given by
Eq \ref{line_CO} is too slow.

Based on detailed analysis of the reactions in a complete 1D model similar to our own, \citet{Moses2011} 
 concluded that the bottleneck
is associated with breaking the C-O bond.  They propose several key channels.
One channel is thermal decomposition of methanol,
\begin{equation}
\label{methanol}
{\rm CH}_3{\rm OH} + {\rm M}  \rightarrow {\rm CH}_3 + {\rm OH} + {\rm M}.
\end{equation}
\citet{Visscher2011} and \citet{Moses2011} state that this is more important when CH$_4$ is abundant.
Another channel, which they state is more important when ${\rm CO} \gg {\rm CH}_4$, is 
\begin{equation}
\label{methylene hydroxide}
{\rm CH}_2{\rm OH} + {\rm H}  \rightarrow {\rm CH}_3 + {\rm OH}.
\end{equation}
The rate of R\ref{methylene hydroxide} depends on the uncertain thermodynamic properties of ${\rm CH}_2{\rm OH}$. 
In the high pressure limit, methanol decomposition R\ref{methanol} is a first order reaction,
\begin{equation}
{\partial \left[{\rm CH}_3{\rm OH}\right] \over \partial t} = -k_{\infty}\left[{\rm CH}_3{\rm OH}\right] .
\end{equation}
Jasper et al (2007) estimate $k_{\infty}$ from theory
\[
k_{\infty} = 6.251 \times 10^{16} \left(300/T\right)^{0.6148} \exp{\left(-46573/T\right)} {\rm ~sec}^{-1}.
\]
 The methanol equilibrium constant 
 \[
 K_{{\rm CH}_3{\rm OH}} = {p{\rm CH}_3{\rm OH}\over p{\rm CO} \cdot p{\rm H}_2^{2}} = 1.1\times 10^{-13}e^{13000/T}
 \]
 relates $\left[{\rm CH}_3{\rm OH}\right]$ to $\left[{\rm CO}\right]$.  The reaction timescale is then 
\begin{eqnarray}
\label{t_mv}
  t_{\rm chem}  & = & \left( k_{\infty}p{\rm H}_2^{2}\, K_{{\rm CH}_3{\rm OH}} \right)^{-1} \nonumber \\
   & = & 4.4\times 10^{-6}\, T^{0.6148} \exp{\left(33573/T\right)}\,p^{-2} f_{{\rm H}_2}^{-2} {\rm ~~sec}.
\end{eqnarray} 
Equation \ref{t_mv} has a weaker dependence on $T$,
 a stronger dependence on $p$ than we find for the ensemble,
 and no dependence on metallicity.

Figure \ref{PB} illustrates quenching using Eq \ref{t_mv} as black triangles.
For the comparison we take $L=0.14\,H$.
This approximates what \citet{Moses2011} may be using.
Equation \ref{t_mv} does well, especially in
cases where ${\rm CO} \ll {\rm CH}_4$ at quenching, which is the regime for which 
\citet{Moses2011} report that Eq \ref{t_mv} applies.
Equation \ref{t_mv} does better if $L\!>\!0.14\,H$.
The match is best for CO with $L=0.2\,H$, and better for CH$_4$ with $L\!\approx\!0.5\, H$.

To reconstruct a simple form for the H$_2$COH channel,
we need to estimate the equilibrium abundance of H$_2$COH.
We use $p{\rm H}_2{\rm COH} = K_{{\rm H}_2{\rm COH}} \cdot p{\rm H}_2 \cdot p{\rm H} \cdot p{\rm CO}$
with $K_{{\rm H}_2{\rm COH}} = 1.0\times 10^{-12}e^{15843/T}$.  
We assume that H and H$_2$ are in equilibrium.
\citet{Jasper2007} list a fast rate, $k=2.8 \times 10^{-10}$ cm$^3$/s, 
for the reaction of H and H$_2$COH.  
Assembling the parts, we obtain
\begin{equation}
\label{moses_CO}
t_{\rm chem} = {2.7\times 10^{15} \exp{\left(38000/T\right)} \over N p  f_{{\rm H}_2}^{-2} } {\rm ~~sec}.
\end{equation}
Results are shown as blue circles on Figure \ref{PB}.
Equation \ref{moses_CO} has a similar $T$ dependence to what we find for the system
as a whole, but the overall rate is about 3 orders of magnitude slower
using $L=0.14\,H$, or about ten times slower using $L=H$.

Disagreement between our model and Moses et al over the H$_2$COH channel is also
apparent in Figure \ref{FF}.
The H$_2$COH radical plays a much more modest role in our 1-D code than it does in \citet{Moses2011}'s.
The same is true for CH$_3$O.
Why this should be so is likely related to differing guesses of H$_2$COH's and CH$_3$O's  
ill-known thermodynamic properties.

There are some more subtle differences between our model and previous quench models
that should be noted.
One is that we find $t_{\rm chem} \propto p^{-1}$ by fitting to the ensemble, 
whilst quench models that attempt to identify the one key forward
reaction tend to get $t_{\rm chem} \propto p^{-2}$.  
This difference is the reason why quench prescriptions in Figure \ref{PB} that do well with 
$f_{\rm CO} \ll f_{{\rm CH}_4}$ miss the mark somewhat with $f_{{\rm CH}_4} \ll f_{\rm CO}$.
Figure \ref{VV} that uses $t_q\propto p^{-1}$ shows fine agreement at both ends of the scale.

Another difference is that we see a well-defined metallicity dependence when fitting to the ensemble.
The metallicity dependence probably stems from the reverse reaction,
which as noted is usually ignored.  
As equilibrium breaks down, the reaction slows in both directions, more quickly in the reverse direction
than in the forward direction.  This behavior is shown clearly in Figure \ref{FF},
where forward, reverse, and net reaction rates are plotted.
It is the net reaction, the difference between the forward and reverse reactions,
that defines the retreat from equilibrium, not the forward reaction alone.
For example, the time scale for the methanol decomposition channel is 
 \begin{equation}
 t^{-1}_{\rm net} = {-1 \over \left[{\rm CO}\right]} {\partial \left[{\rm CO}\right] \over \partial t} = { k_{f} \left[{\rm CH}_3{\rm OH}\right] \over \left[{\rm CO}\right] } - { k_{r} \left[{\rm CH}_3\right] \left[{\rm OH}\right] \over \left[{\rm CO}\right] }.  
 \end{equation} 
The reverse reaction is quadratic in metallicity, whilst [CO] is quadratic in metallicity
only for $f_{\rm CO} \ll f_{{\rm CH}_4}$, which suggests that it is through the reverse reaction
that metallicity enters $t_{\rm CO}$.  
The same consideration applies for all the important reactions in Figure \ref{FF}, as the reverse reaction in
every case is between CH$_3$ and OH.

 \section{Quenching in the N$_2$-NH$_3$ system}

\begin{figure}[!htb] 
   \centering 
\includegraphics[width=.45\textwidth]{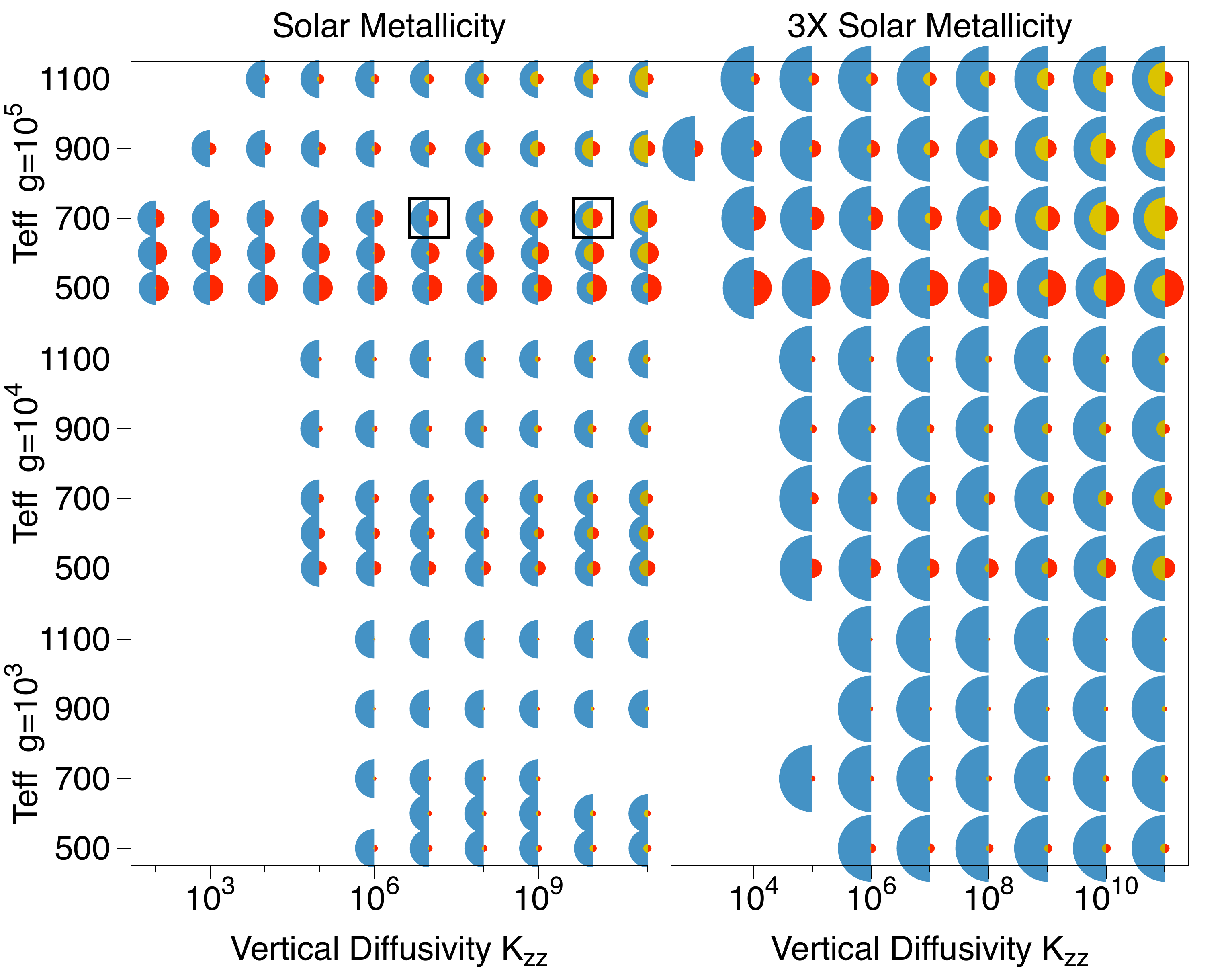} 
   \caption{\small Nitrogen. The six panels are as in Fig. \ref{ensemble}.
   Relative abundances of NH$_3$ (red, east) and N$_2$ (blue, west) are indicated by area.
   HCN abundances are multiplied by 5 to make them visible at lower $g$ 
   and superposed in gold on N$_2$. 
   Higher gravities and cooler temperatures favor NH$_3$.
   The NH$_3$/N$_2$ ratio is insensitive to $K_{zz}$ and only weakly sensitive to $m$,
   which suggests that NH$_3$ can be a proxy for $g$. 
   HCN fares best at high $g$ and high $K_{zz}$.
   The HCN/NH$_3$ ratio is sensitive to $K_{zz}$ and only weakly sensitive to metallicity.
   The two boxed models are ilustrated in Figure \ref{reviewer_added}.}
\label{ZZ}
\end{figure}

The approach we used to search for quenching of CO and CH$_4$ does not work well for N$_2$ and NH$_3$. 
Indeed, Figure \ref{ZZ}---the analog to Fig \ref{ensemble} for N$_2$ and NH$_3$---shows no
sign that quenching plays {\em any} role in the N$_2$-NH$_3$ reaction, although doubtless quenching occurs.  

Two things get in the way.  Of less importance, a third species---HCN---becomes non-negligible at high temperatures.
HCN also quenches at high temperature, which leaves
the nitrogen system with three possibly distinct quench points:  NH$_3$-N$_2$, NH$_3$-HCN, and N$_2$-HCN.
Although HCN is never very abundant in equilibrium, it is often abundant enough that its decomposition
can increase the NH$_3$ abundance by more than 10\% after the NH$_3$-N$_2$ reaction has quenched. 
When confused with the second, greater obstacle, HCN can make it difficult to pinpoint where quenching occurs.

\begin{figure*}[!htb] 
   \centering 
\includegraphics[width=0.8\textwidth]{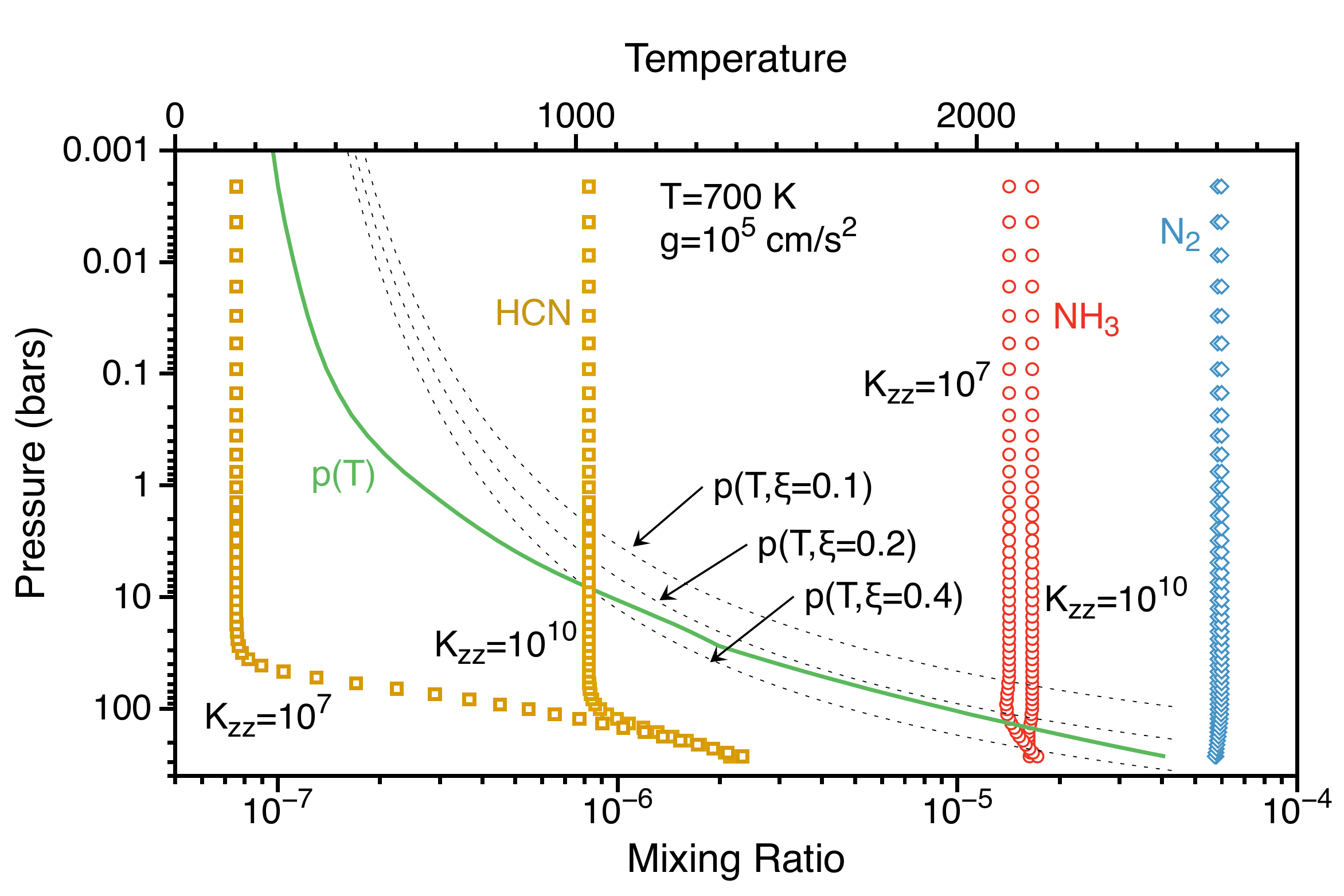} 
   \caption{\small Two examples of quenching in nitrogen.
   The particular models are indicated by boxes on Figure \ref{ZZ}.
   Volume mixing ratios of N$_2$, NH$_3$, and HCN are shown for two different $K_{zz}$ pertinent
   to brown dwarfs.   
   The relative insensitivity of NH$_3$ and N$_2$ to $K_{zz}$ is a consequence of contours of
   constant ${\rm NH}_3/{\rm N}_2$ being nearly parallel to the adiabat $p(T)$.
   This is illustrated here by showing three relevant contours of constant $\xi \equiv {\rm NH}_3/{\rm N}_2$.
   Both $\xi$ and $p(T)$ are plotted against the temperature axis (top). 
   In the full models $\xi = 0.28$ and $\xi = 0.24$ for $K_{zz}=10^{10}$ cm$^2$/s and $K_{zz}=10^{7}$ cm$^2$/s, respectively.
}
\label{reviewer_added}
\end{figure*}

The greater obstacle is that curves of constant NH$_3$/N$_2$ are nearly parallel to the
adiabat at the pressures and temperatures where quenching occurs.  
This is illustated in Figure \ref{reviewer_added}.
Consequently the NH$_3$/N$_2$ ratio in a parcel remains close to equilibrium well after quenching has taken place,
which makes the NH$_3$/N$_2$ ratio indifferent to quenching \citep{Saumon2006}.
To show this, write the N$_2$-NH$_3$ equilibrium in the form
\begin{equation}
\label{NH3_equilibrium}
K_{{\rm NH}_3 \cdot {\rm N}_2} = 5.90\times 10^{-13}\exp{\left(13207/T\right)} = { p{\rm NH}_3^2 \over p{\rm N}_2 \!\cdot\! p{\rm H}_2^3 } .
\end{equation}
Define $\xi \equiv {\rm NH}_3/{\rm N}_2$. 
If we approximate the total mixing ratio of N by $f_{\rm N} \approx f_{{\rm NH}_3} + 2f_{{\rm N}_2}$ (including HCN makes this very complicated),
we can write 
\begin{equation}
\label{KNH3_2}
\xi^2 \left( {f_{\rm N} \over 2+\xi } \right) = p^2 f_{{\rm H}_2}^3 A e^{B/T}.
\end{equation}
With context-obvious substitutions, the temperature gradient for contours of constant $\xi$ is
\begin{equation}
\label{KNH3_3}
\left({dT\over dp}\right)_{\xi} = {2T\over B} \, {T \over p}  \approx 0.23 {T \over 1500} \, {T \over p} .
\end{equation}
The temperature gradient in Eq \ref{KNH3_3} is parallel to the adiabat (Eq \ref{adiabatic_gradient})
at 1750 K, and nearly parallel for $1500 < T < 2000$ K.
For temperatures initially greater than 1750 K, the equilibrium ratio of NH$_3$/N$_2$ decreases as the parcel cools,
reaching a minimum at $\sim$1750 K where $\left({dT/dp}\right)_{\xi} = \left({dT/dp}\right)_{S}$.
If the parcel remains in chemical equilibrium, NH$_3$/N$_2$ will increase again as it cools further.
This then is how we explain the insensitivity of 
NH$_3$/N$_2$ to $K_{zz}$ seen in Figure \ref{ZZ}:  the NH$_3$/N$_2$ ratio computed by the full kinetics model
is near the equilibrium value at the temperature where contours of NH$_3$/N$_2$ are parallel to the adiabat,
which is also the minimum equilibrium value in the atmosphere.
For the kinds of atmospheres considered in this paper, it appears that 
the amount of ammonia in the visible parts of the atmosphere
will be comparable to the minimum equilibrium abundance computed along the adiabat.

On the other hand, because our results are insensitive to quenching, it is not difficult to find a non-unique chemical reaction time scale 
for the nitrogen system that predicts $f_{{\rm NH}_3}$ well.
A chemical reaction time scale that works well in a quenching scheme that sets the reaction rate time $t_{{\rm NH}_3}$ equal to the mixing
time $t_{\rm mix} = H^2/K_{zz}$ is
\begin{equation}
\label{t_N} 
 t_{{\rm NH}_3} = 1.0\times 10^{-7} p^{-1} \exp{\left(52000/T\right)} {\rm ~sec.}  
\end{equation}
This particular choice presumes that the energy barrier is set by ${\rm N}_2 + {\rm H}_2 \rightarrow {\rm NNH} + {\rm H}$.
This choice of $t_{{\rm NH}_3}$ is not unique. 
 Almost any plausible choice of Arrhenius parameters
 that gives roughly the same time scale as Eq \ref{t_N} at 10 bars and 1750 K will work just as well. 
We have no information to constrain dependence on $p$ or $m$.

\begin{figure*}[!htb]
\plottwo{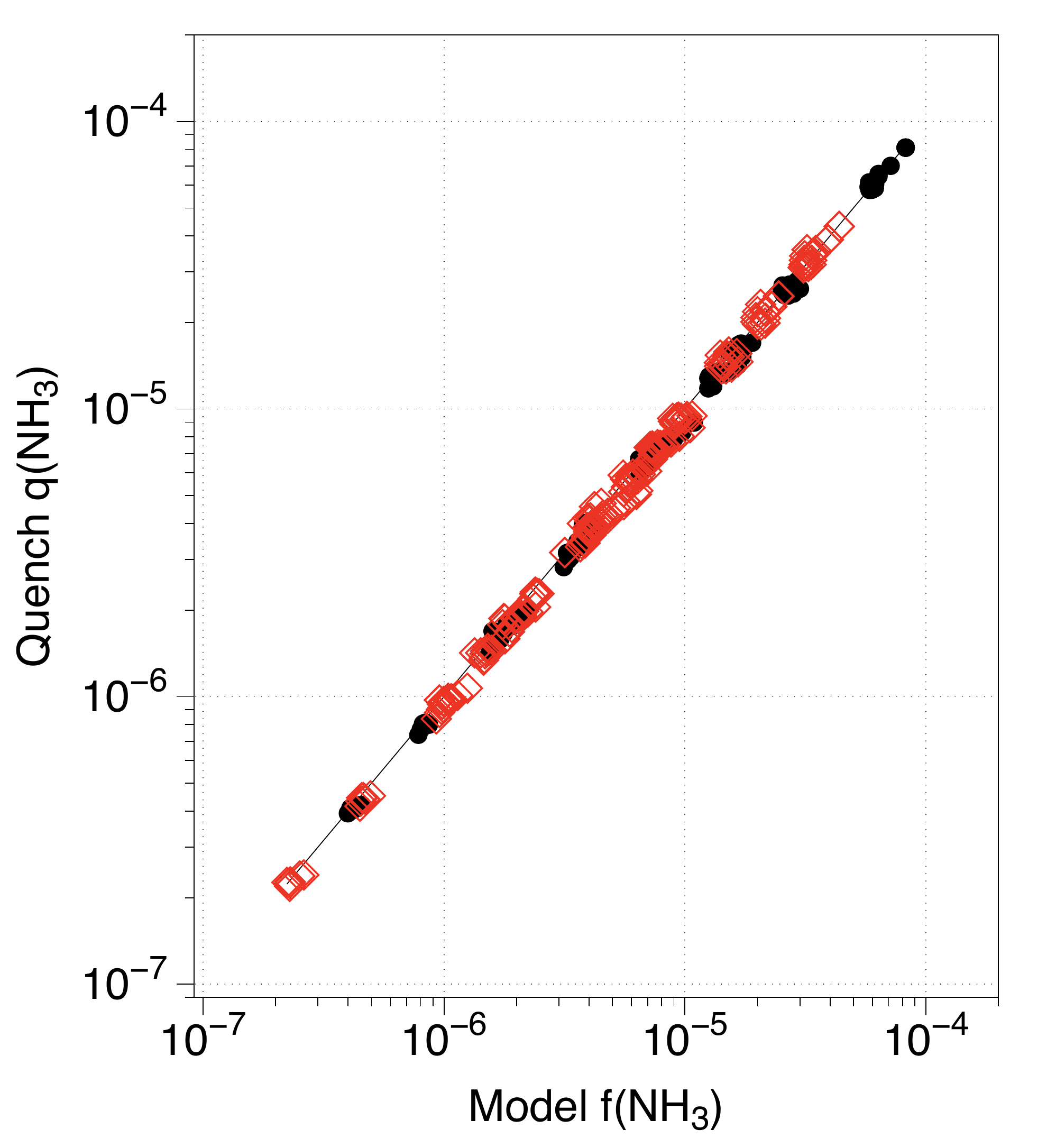}{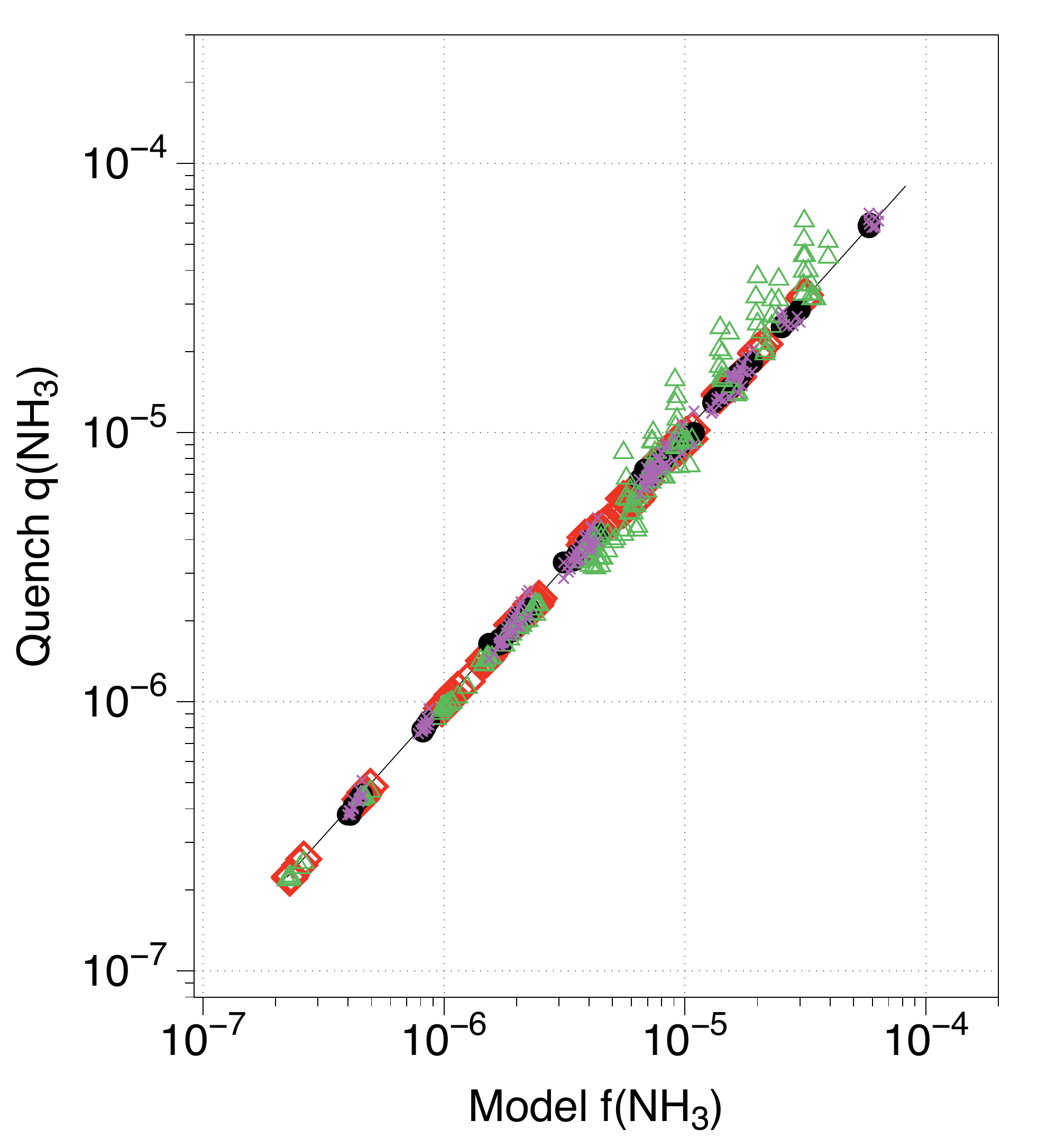}
  \caption{\small Ammonia predicted by the quenching approximation $q({\rm NH}_3)$
 (y-axes)
 plotted against the
   actual
  asymptotic mixing ratio $f({\rm NH}_3)$ determined from the ensemble of 1-D kinetics models
    (x-axes). A perfect approximation would adhere to the line.
   Open red symbols are for $m\!=\!1$ and solid black symbols are for $m\!=\!3$. 
{\it Left.} Using Eq \ref{t_N}.
{\it Right.} 
   Open red diamonds $m\!=\!1$ and filled black circles $m\!=\!3$ are based on \citet{Lodders2002};
   open green triangles and purple crosses are based on \citet{Line2011}.
   }
   \label{QQ}
\end{figure*}

Figure \ref{QQ} shows how our expression compares to some expressions in the literature.
The chemical reaction time recommended by \citet{Lodders2002} is equivalent to  
\begin{equation}
\label{k_fl}
t_{\rm chem} = { 1.2\times 10^{8} \exp{\left(81515/T\right)} \over p \!\cdot\! N \!\cdot\! f_{{\rm H}_2}^2} {\rm ~~sec}.
\end{equation}
Equation \ref{k_fl} is based on the possible reaction ${\rm N}_2 +{\rm H}_2 \rightarrow {\rm NH}_2 + {\rm N}$.
Figure \ref{QQ} shows that Eq \ref{k_fl} used with $L=H$ gives as good a fit
to our models as Eq \ref{t_N}. 
The extreme temperature dependence of Eq \ref{k_fl} appears justified.

\citet{Line2011} look for quenching in the reaction ${\rm H}_2 +  {\rm N}_2{\rm H}_2 \rightarrow 2{\rm NH}_2$.
Using rates given by \citet{Line2011}, and using
$p{\rm N}_2{\rm H}_2 = K_{{\rm N}_2{\rm H}_2}\!\cdot\! p{\rm N}_2\!\cdot\! p{\rm H}_2$ with
$K_{{\rm N}_2{\rm H}_2} = 3.8\times 10^{-6} e^{-25738/T}$, we get
\begin{equation}
\label{line}
t_{\rm chem} = {1.3\times 10^{11}\, T^{0.93}\exp{\left(46400/T\right)} \over N \, p \, f_{{\rm H}_2}^2} {\rm ~~sec}
\end{equation}
as our best effort to reproduce their chemical time scale for NH$_3$ equilibration with N$_2$.
Figure \ref{QQ} shows that Eq \ref{line} agrees well with the predictions of our kinetics model when $m\!=\!3$,
but sometimes predicts more NH$_3$ than we find for $m\!=\!1$.
Disagreement is limited to cases with $K_{zz} < 10^7$ cm$^2$/s, 
which means that Eq \ref{line} is relatively fast at the lowest quench temperatures. 
It is possible that the temperature dependence of Eq \ref{line} is not steep enough,
or that the bottleneck involves N$_2$ rather than N$_2$H$_2$, or that the highly uncertain
thermodynamic parameters of NNH are being treated differently between models.

\subsection{HCN-NH$_3$-N$_2$}
The approach we used for CO and CH$_4$ works moderately well for HCN and NH$_3$
and less well for HCN and N$_2$. 

The equilibrium between HCN and CH$_4$ and NH$_3$,
\begin{equation}
\label{HCN_eq1}
K_{{\rm HCN} \cdot {\rm CH}_4} = { p{\rm CH}_4\!\cdot\! p{\rm NH}_3 \over p{\rm HCN} \!\cdot\! p{\rm H}_2^3 } = 3.0\times 10^{-14}\exp{\left(33460/T\right)} ,
\end{equation}
closely resembles the parallel equilibrium Eq \ref{K_eq} between CO and CH$_4$.
For the cool objects in which CH$_4$ is abundant at depth, the strong temperature dependence of $K_{{\rm HCN}\cdot {\rm CH}_4}$ means that HCN fares best
with respect to NH$_3$ at high temperature when parcels move up or down along an adiabat.

For warmer worlds where CO and N$_2$ are dominant at depth, the most informative equilibrium is with CO, N$_2$, H$_2$, and H$_2$O, all of which are nearly constant when nearly all the C is in CO.
The formal reaction is
$2{\rm CO} + {\rm N}_2 + 3{\rm H}_2 \leftrightarrow 2{\rm HCN} + 2{\rm H}_2{\rm O}$,
and the corresponding equilibrium constant is 
\begin{equation}
\label{HCN_eq2}
K_{{\rm HCN} \cdot {\rm N}_2} = { p{\rm CO}^2 \!\cdot\! p{\rm N}_2 \!\cdot\! p{\rm H}_2^3 \over p{\rm HCN}^2 \!\cdot\! p{\rm H}_2{\rm O}^2 } = 4.278\times 10^{11} \exp{\left(-528/T\right)} ,
\end{equation}
for which the temperature dependence is weak at quenching where $T_q\approx 2000$ K. 
Equation \ref{HCN_eq2} indicates that $f_{\rm HCN} \propto p$.
Hence, in both the cool and the warm limits HCN increases with depth in deep atmospheres.
Thus to identify HCN quenching it suffices to start from the bottom and find the altitude where the HCN abundance stops decreasing. 

\begin{figure*}[!htb]
\plottwo{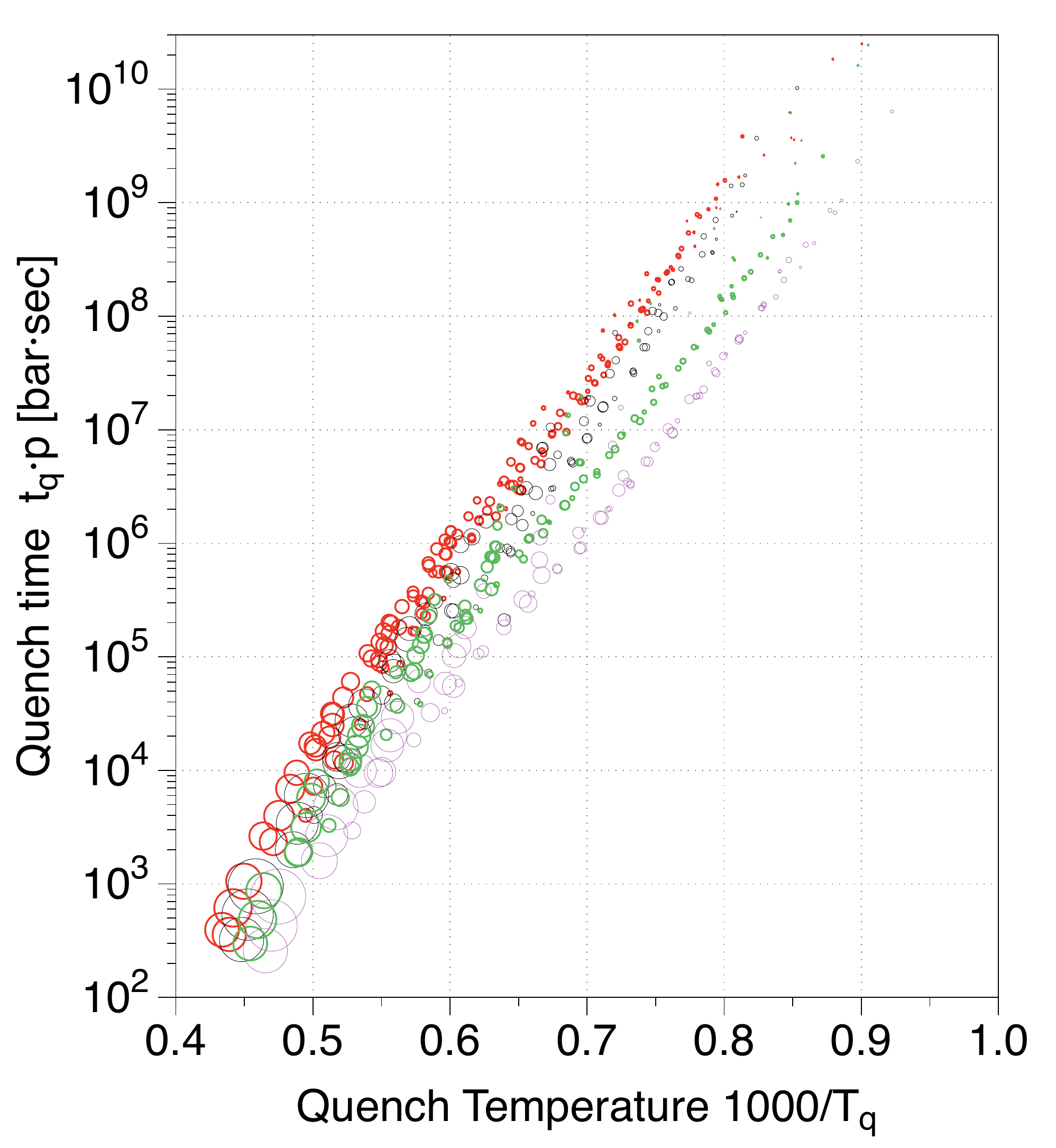}{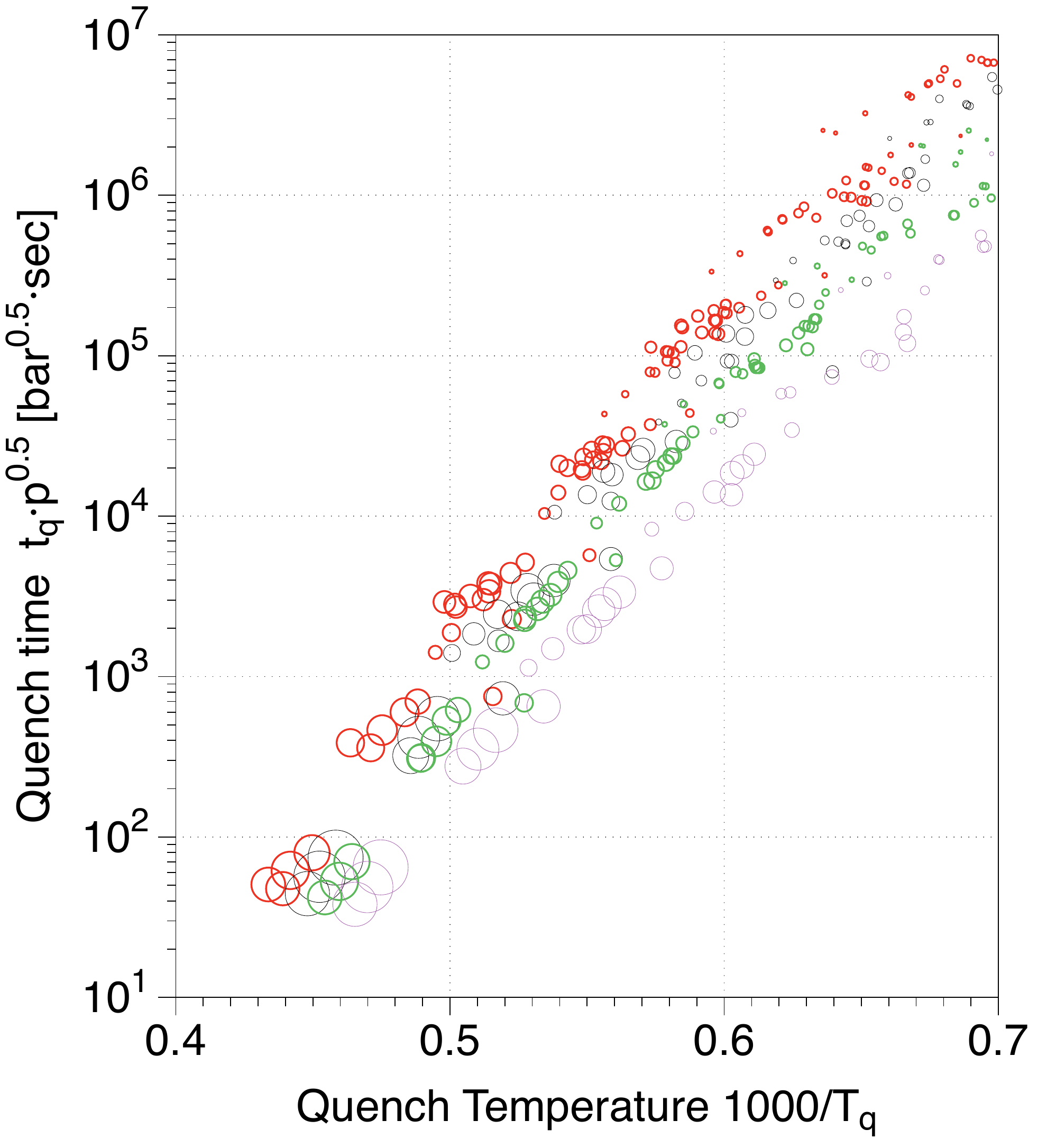}
  \caption{\small HCN quenching time scales determined from the ensemble of models
   with respect to NH$_3$ and CH$_4$ (red $m\!=\!1$ and black $m\!=\!3$) and CO (green $m\!=\!1$ and violet $m\!=\!3$).
   Symbol areas are proportional to $f_{\rm HCN}$. 
{\it Left.} The time scale plotted is $t_q \propto p_q^{-1}$.
{\it Right.} 
   High $T_q$ results fit better to the Arrhenius form with time scale $t_q \propto p_q^{-0.5}$.
   }
\label{CC}
\end{figure*}

We looked for quench points defined against Eq \ref{HCN_eq1}, Eq \ref{HCN_eq2}, and a third equilibrium with CO and NH$_3$,
\begin{equation}
\label{HCN_eq3}
K_{{\rm HCN} \cdot {\rm CO}} = { p{\rm CO} \!\cdot\! p{\rm NH}_3  \over p{\rm HCN} \!\cdot\! p{\rm H}_2{\rm O} } = 0.5025 \exp{\left(6339/T\right)} .
\end{equation}
Quenching with N$_2$ ($K_{{\rm HCN} \cdot {\rm N}_2}$) gives an indifferent fit to an Arrhenius relation, as might be expected
given the insensitivity of Eq \ref{HCN_eq2} to $f_{\rm HCN}$ and the high thermal stability of N$_2$.
The other two equilibria both give plausible Arrhenius-like fits, although far from perfect (Figure \ref{CC}).

A direct fit to the chemical time scale derived from the equilibrium $K_{{\rm HCN} \cdot {\rm CH}_4}$
 for the full ensemble of models is
\begin{equation}
\label{HCN_q1} 
 t'_{\rm q1} = 1.6\times 10^{-4} p^{-1} m^{-0.7} \exp{\left(37000/T\right)} {\rm ~~sec,}  
\end{equation}
whilst the corresponding fit to $K_{{\rm HCN} \cdot {\rm CO}}$ is
\begin{equation}
\label{HCN_q2} 
 t'_{\rm q2} = 1.3\times 10^{-4} p^{-1} m^{-1} \exp{\left(34500/T\right)} {\rm ~~sec.}  
\end{equation}
At quench temperatures greater than 1600 K, the pressure dependence is better described by $t'_{\rm q} \propto p^{-0.5}$,
and the $T$ dependence is stronger with an Arrhenius $B$-factor of order 46000 K.
The stronger temperature dependence for high $T_q$  suggests that reactions with N$_2$ with its high activation energy are becoming important.
The higher $T_q$ cases correspond to higher $f_{\rm HCN}$. 

\begin{figure*}[!htb]
\plottwo{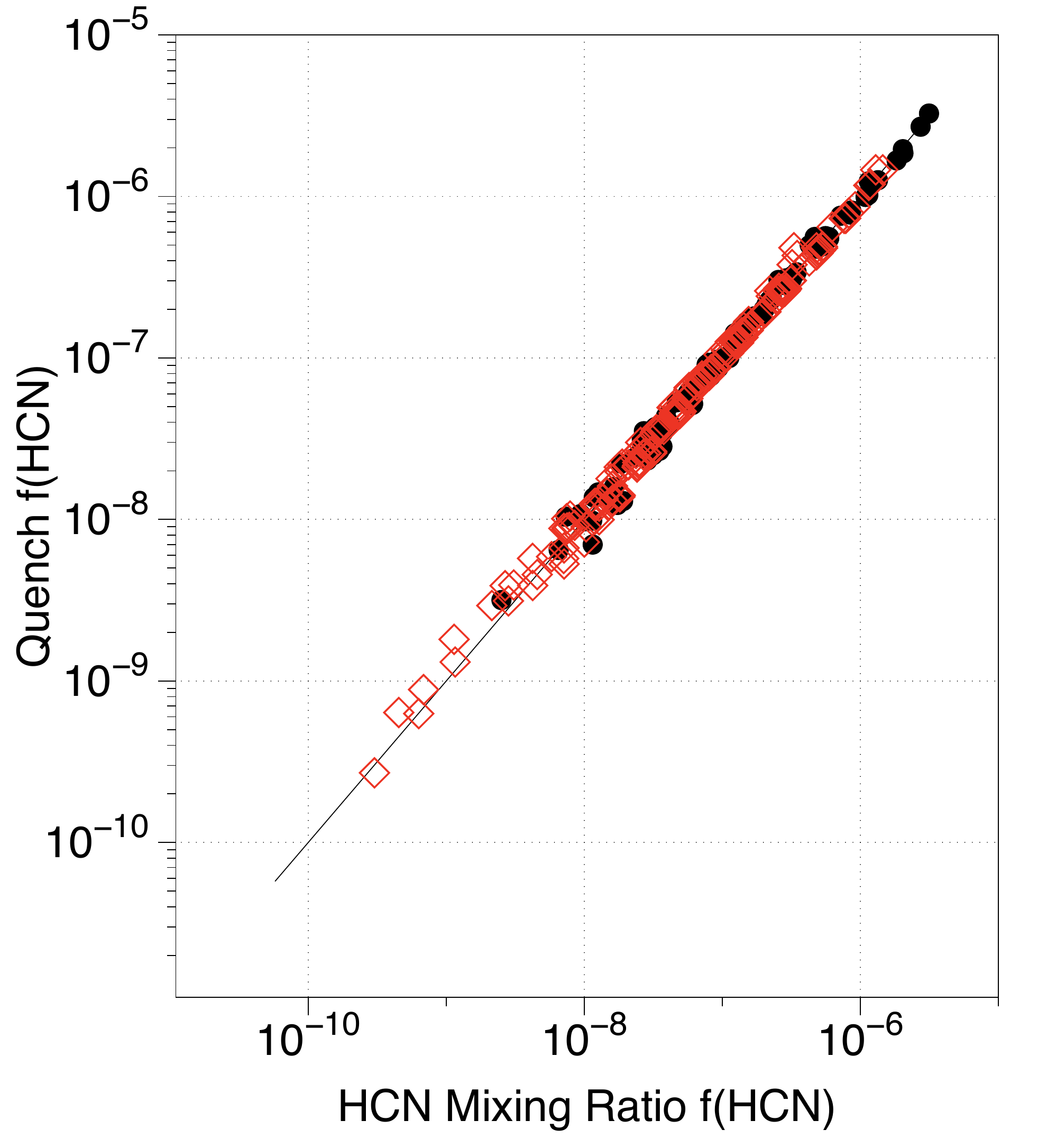}{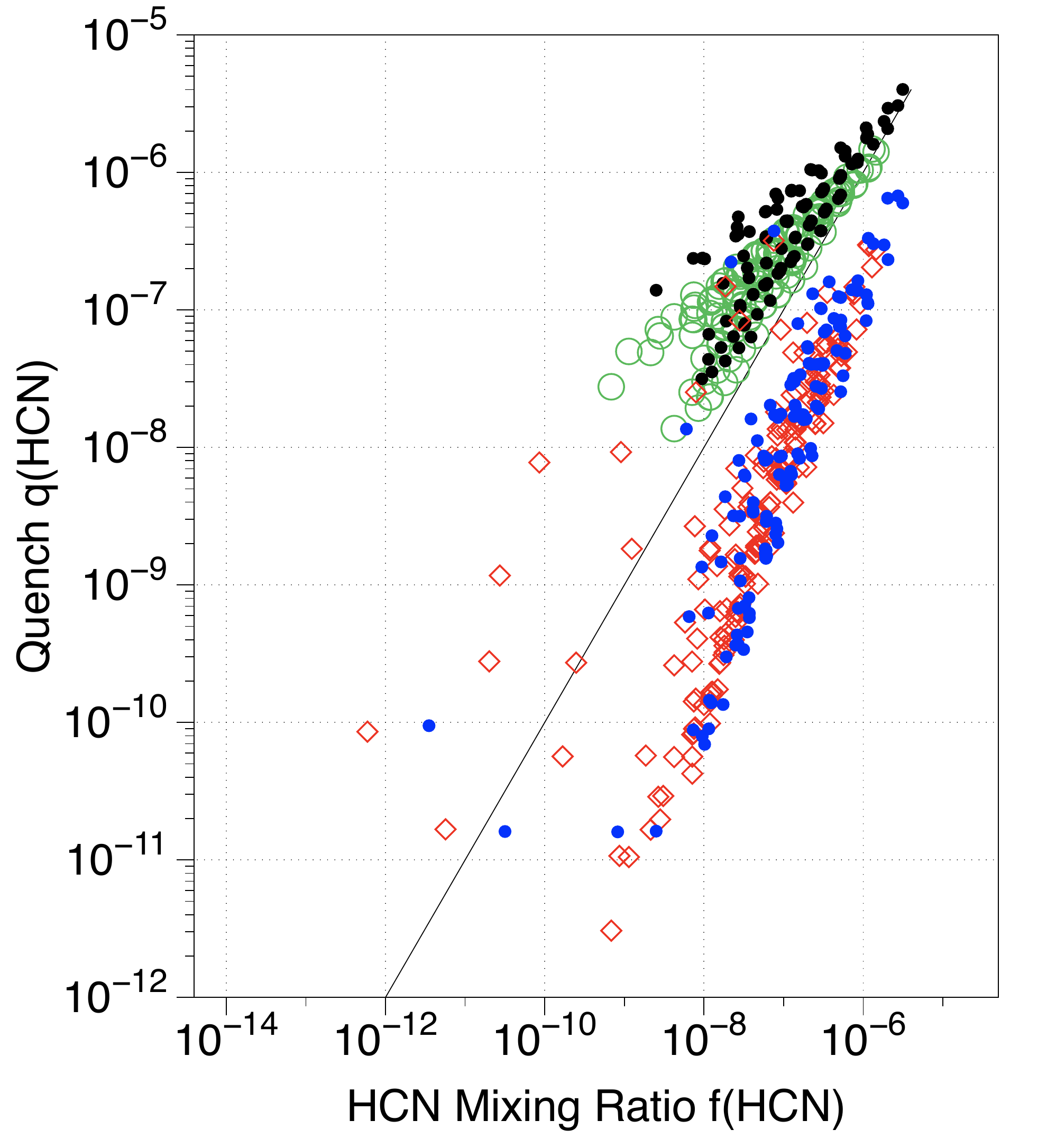} 
  \caption{\small HCN mixing ratios predicted by various quenching
approximations (y-axes)
plotted against the 
actual
asymptotic HCN mixing ratio as computed by the full ensemble of 1-D kinetics models
(x-axes).  A perfect approximation would adhere to the line.
{\it Left.} Quench predictions using Eq \ref{HCN_quench}. 
   Open red symbols are for $m\!=\!1$ and solid black symbols are for $m\!=\!3$. 
{\it Right.} 
   Quench approximations using Eq \ref{Fegley_HCN} with $L=H$
   (open green circles, $m\!=\!1$; filled black circles, $m\!=\!3$)
   and using Eq \ref{moses_HCN} with $L=0.14\,H$
   (open red diamonds, $m\!=\!1$; filled blue circles, $m\!=\!3$). 
   }
   \label{EE}
\end{figure*}

As noted above, the quench approximation is not very sensitive to the details of $t_q$, and that is the case here as well.
Computed quenched abundances using $t_q=t_{\rm mix}=H^2/K_{zz}$ provide a good approximation to the HCN mixing ratios computed
by the full model with
\begin{equation}
\label{HCN_quench} 
 t_{\rm HCN} = 1.5\times 10^{-4} p^{-1} m^{-0.7} \exp{\left(36000/T\right)}  {\rm ~sec.}  
\end{equation}
This expression seems to work well for all cases we have considered (Fig \ref{EE}).

\citet{Fegley1996} treat HCN destruction as controlled by direct reaction of HCN with H$_2$ to make NH and CH$_2$.
The corresponding chemical time scale is
\begin{equation}
\label{Fegley_HCN}
t_{\rm chem} = { 9.3\times 10^7 \exp{\left(70456/T\right)} \over  N f_{{\rm H}_2}^{2} } {\rm ~sec.}
\end{equation}
This has a very steep temperature dependence.
Quench approximations using Eq \ref{Fegley_HCN} with $L=H$ are shown in Figure \ref{EE}.
This approximation predicts much more HCN 
(a higher quench temperature) than we find in our 1-D models, a result consistent with
the steep temperature dependence of Eq \ref{Fegley_HCN}.

\citet{Moses2010} presume that HCN destruction is controlled by reaction of H$_2$ with the H$_2$CN radical.
This is a much faster reaction.
To convert their discussion into a reaction time requires defining the H$_2$CN equilibrium abundance.
We write $p{\rm H}_2{\rm CN} = K_{{\rm H}_2{\rm CN}}\!\cdot\! p{\rm HCN} \!\cdot\! p{\rm H}$ with
$K_{{\rm H}_2{\rm CN}} = 1.0\times 10^{-6} e^{14240/T}$.
This is likely not the same as what \citet{Moses2010} use.
Atomic and molecular hydrogen are also assumed to be in equilibrium. 
Other pertinent information is given in \citet{Moses2010}.  
The reaction time scale that results is
\begin{equation}
\label{moses_HCN}
t_{\rm chem} = {8.3\times 10^{20} \exp{\left(23358/T\right)} \over T^{1.941} N p^{0.5} f_{{\rm H}_2}^{1.5} } {\rm ~~sec.}
\end{equation} 
Figure \ref{EE} shows that Eq \ref{moses_HCN} used in a quench approximation
predicts much less HCN than we compute in our 1D models.
This means that reactions destroying HCN are occurring at relatively low temperatures.
This fits with the relatively weak temperature dependence of $t_{\rm chem}$
in Eq \ref{moses_HCN}.

The comparison of models may be frustrated in part by a hole in our model.
We did not include methylamine (CH$_3$NH$_2$), which \citet{Moses2010}
argue plays the same role in hydrogenation of HCN at low temperatures and high pressures 
that methanol plays for CO.
Their scheme is plausible but almost entirely hypothetical because it passes through several free radicals 
that must exist but about which little else is known. 
Our omission of a CH$_3$NH$_2$ channel implies that Eq \ref{HCN_quench} overestimates HCN,
especially in cool worlds.
Another issue undermining comparison between models is that we have not implemented \citet{Moses2010}'s 
full quench scheme: \citet{Moses2010} require that HCN quench with respect to already
quenched abundances of CH$_4$ and NH$_3$.  
But at the temperatures at which HCN might actually be 
abundant enough to be detectable in EGPs and BDs,
our prescription should work and is easy to use.

\section{CO$_2$}

In principle CO$_2$ is also subject to quenching \citep{Prinn1987}, with caveats.
First, because CO$_2$ quenches at a lower temperature than CO,
it quenches with respect to the disequilibrium (quenched) abundance of CO.
Second, in practice, CO$_2$ can be much enhanced by photochemistry if
the world in question is subject to significant stellar irradiation.
Under such conditions quenching is a poor guide.
But for solitary brown dwarfs and planets in wide orbits it should do fine.

\begin{figure*}[!htb]
\plottwo{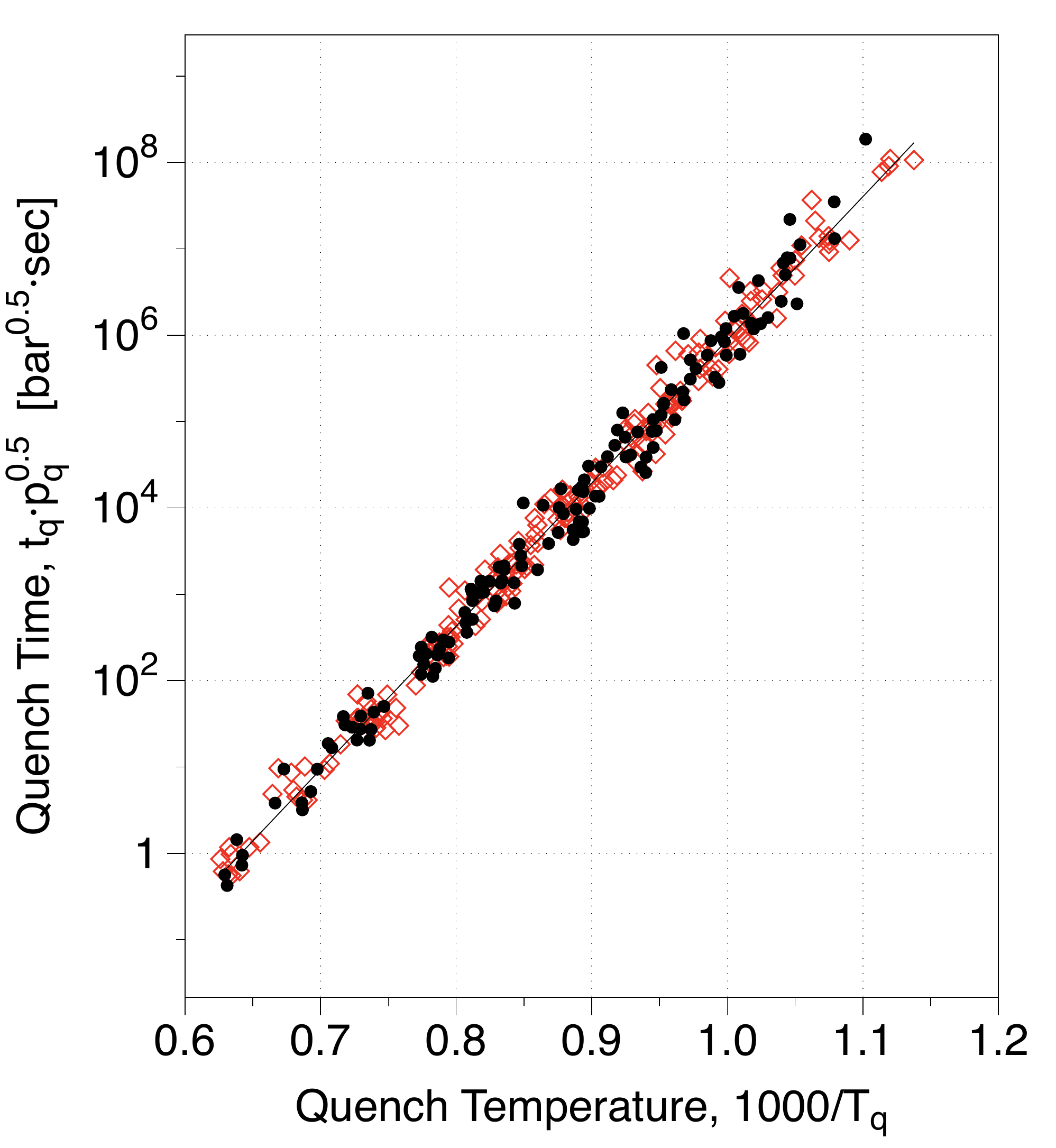} {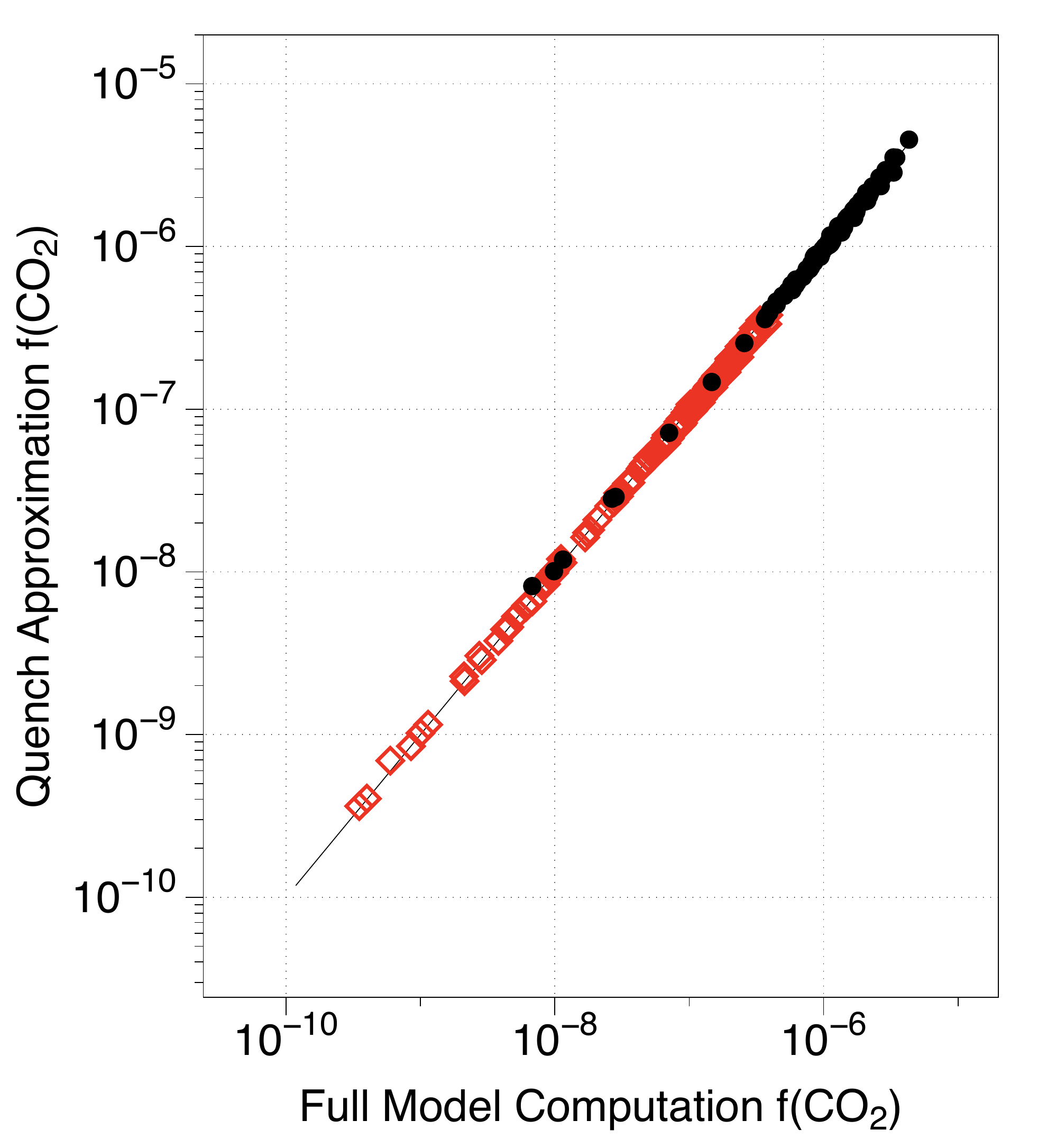}
  \caption{ \small
{\it Left.} CO$_2$ quenching time scales determined from the ensemble of models.
Here quenching refers to disequilibrium between CO$_2$ and CO and H$_2$O;
   CO and H$_2$O are themselves not in equilibrium because they have already quenched.
   The time scale plotted is $t_q \times p_q^{0.5}$.
{\it Right.} $f_{{\rm CO}_2}$ predicted by the quenching approximation 
(y-axis)  plotted against the  actual
   asymptotic mixing ratio determined from the full ensemble of 1-D kinetics model
  (x-axis). A perfect approximation would adhere to the line.
   Open red symbols are for $m=1$ and solid black symbols are for $m\!=\!3$.
   }    
\label{CO2}
\end{figure*}

The approach is similar to that used for CO and CH$_4$ above.
Equilibrium between CO$_2$, CO, H$_2$, and H$_2$O can be approximated by
\begin{equation}
\label{KCO2}
K_{{\rm CO}_2} = {p{\rm CO}\!\cdot\! p{\rm H}_2{\rm O} \over p{\rm CO}_2\!\cdot\! p{\rm H}_2} = 18.3\exp{\left(-2376/T-\left(932/T\right)^2 \right)} .
\end{equation}
The equilibrium product Eq \ref{KCO2} is evaluated using the quenched values of $p{\rm CO}$ and $p{\rm H}_2{\rm O}$,
beginning at the altitude where CO and CH$_4$ quench, and then extending to all higher altitudes. 
CO$_2$ quenching is pinned at the altitude where the equilibrium defined by Eq \ref{KCO2} breaks down.
Figure \ref{CO2} shows the results of doing so for the ensemble of models.
Figure \ref{CO2} is noisy because the deviations from equilibrium are modest and can go in either direction.
It is interesting that the quenching timescale $t_q$ varies inversely with the square root of the quench pressure $p_q$,
and that unlike the CO-CH$_4$ system there is no discernible dependence on metallicity.
The temperature dependence in the Arrhenius-like relation $t_q\cdot p^{0.5}_q \propto \exp{\left(38000/T\right)}$ is similar to the other cases
we have looked at involving CO, which is also notable.

Figure \ref{CO2} shows that CO$_2$ abundances in the ensemble of models are well approximated by a quench model provided
that the appropriate disequilibrium CO and H$_2$O mixing ratios are used.
A chemical reaction timescale that works well for CO$_2$ quenching is
\begin{equation}
\label{t_CO2}
 t_{{\rm CO}_2} = 1.0\times 10^{-10} \, p^{-0.5} \exp{\left(38000/T\right)} {\rm ~~sec} 
\end{equation}
where $p$ is in bars.  
The results shown in Fig \ref{CO2} are rather insensitive to the Arrhenius $A$ factor in
$t_{{\rm CO}_2}$, which perhaps is to be expected given the weak temperature dependence of the 
equilibrium constant Eq \ref{KCO2} compared to the very strong temperature dependence of Eq \ref{t_CO2}.

\section{Detectability}
\begin{figure}[!htb] 
   \centering
\includegraphics[width=.45\textwidth]{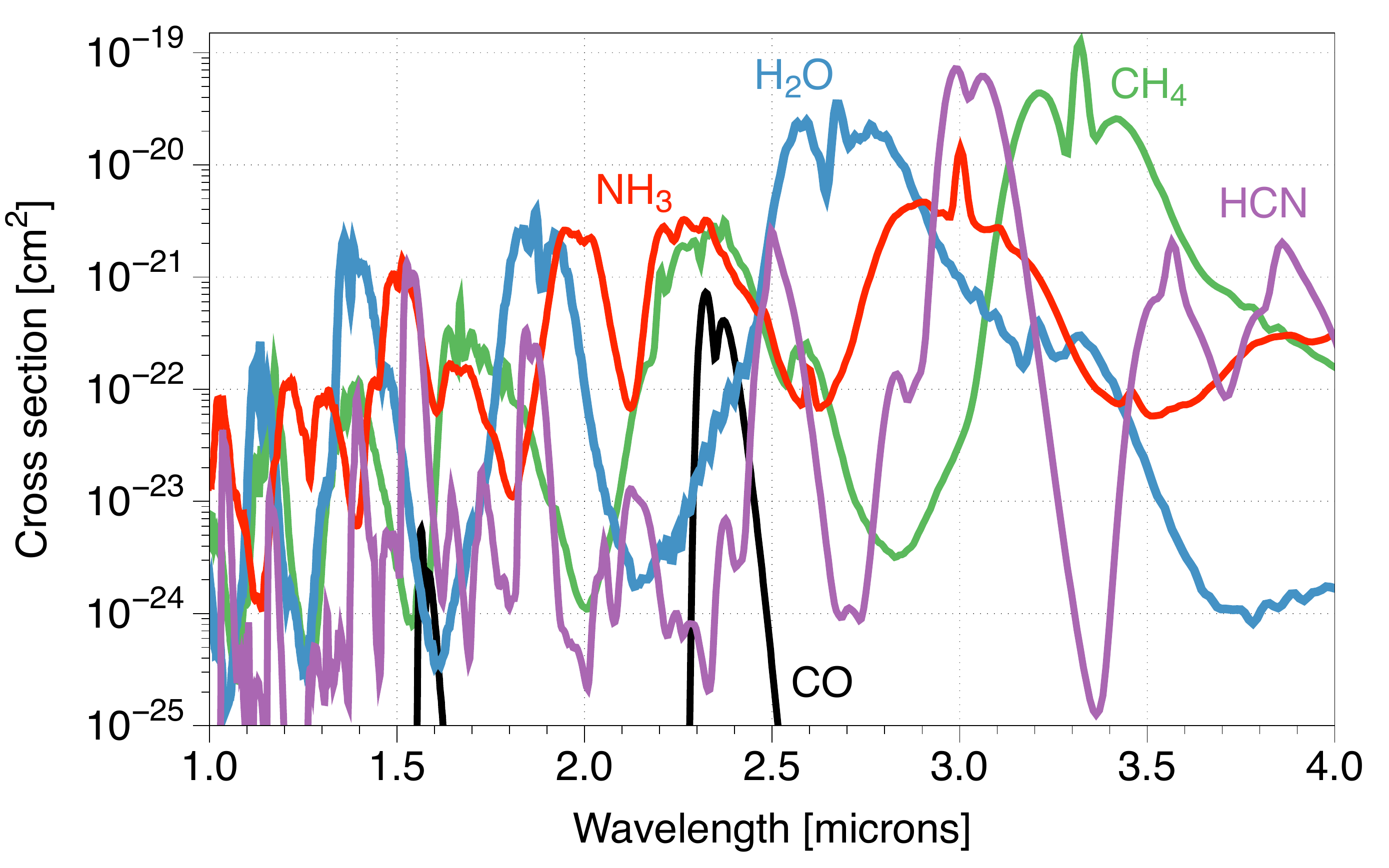} 
   \caption{\small Absorption cross sections of molecules of interest here, computed at 650 K and 1 bar pressure, 
  for wavelengths between 1 and 4 microns. 
 UCL opacities are provided by the University College London and the Hitemp opacities are from the HITRAN opacity database \citep{Rothman2010}. }
\label{Freedman}
\end{figure}

\begin{figure*}[!htb] 
\plottwo{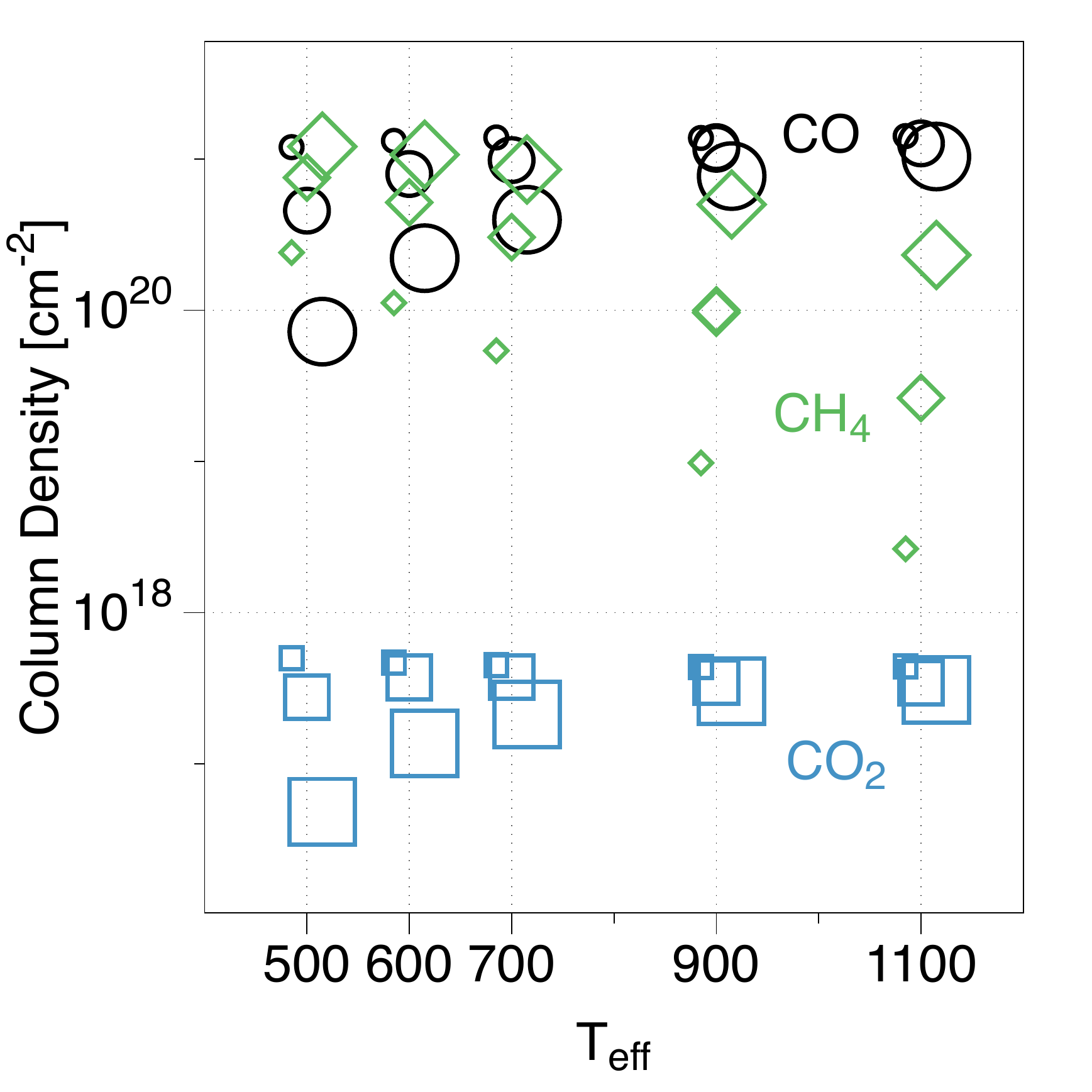}{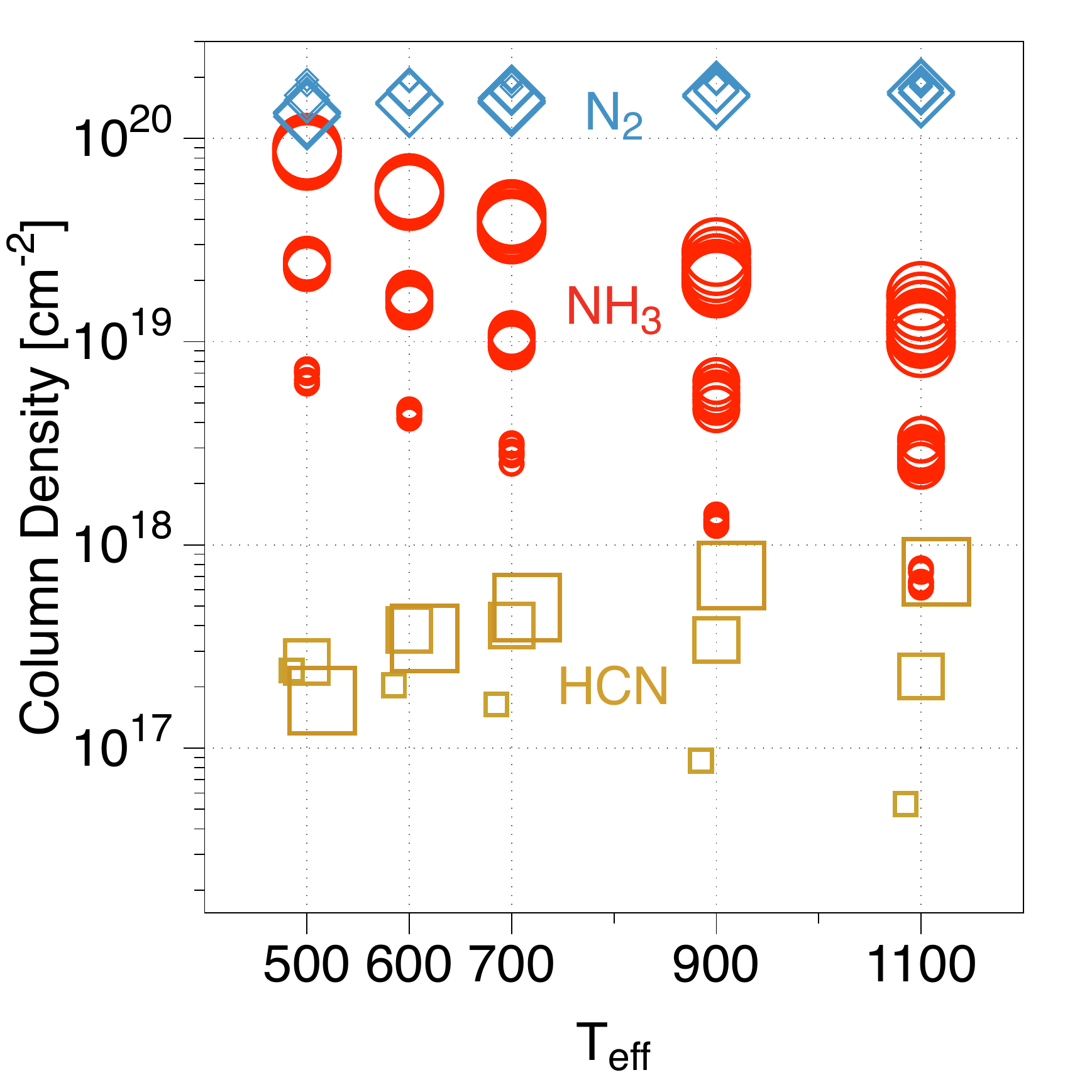} 
  \caption{\small A sequence of models that illustrates how column densities of important carbon-bearing species
  change with temperature $T_{\rm eff}$ and gravity $g$. 
  $K_{zz}$ is restricted to $10^{13}/g$ for CO, CH$_4$, CO$_2$, HCN.
  All $K_{zz}$ are plotted for N$_2$ and NH$_3$.
  Columns are integrated above $p=10^{-5}\!\cdot\!g$ bars. 
   Small symbols represent $g\!=1000$, mid-sized symbols $g\!=\!10^4$, and big symbols $g\!=\!10^5$.
 {\em Left.} CO (black circles), CH$_4$ (green diamonds), CO$_2$ (blue squares).
 {\em Right.} N$_2$ (blue diamonds), NH$_3$ (red circles), HCN (gold squares).
}
\label{columns}
\end{figure*}

Whether or not a molecule can be detected depends on the abundance and opacity of the species in question and on the opacities of other molecules and clouds.  
For reference, Figure \ref{Freedman} shows absorption cross sections at 650 K and 1 bar pressure for ${\rm H}_2{\rm O}$, $\rm NH_3$, $\rm CH_4$, HCN, and CO.
These can be compared to illustrative column densities shown in Figure \ref{columns}.
The latter are integrated upward from the 1, 0.1, and 0.01 bar pressure level,
  typical near-IR photospheric pressures for planets or brown dwarfs with gravities of $10^5$, $10^4$, and $10^3\,\rm cm\,s^{-2}$, respectively. 
For CO, CH$_4$, CO$_2$, and HCN we plot column densities for only one value of $K_{zz}$ for each $(g,T_{\rm eff})$ pair. 
We arbitrarily select $K_{zz}=10^{13}/g$, a high value of $K_{zz}$ but consistent with Eq \ref{Gierasch}, for the illustration.
For N$_2$ and NH$_3$ we plot column densities for all $K_{zz}$ to emphasize how little these depend on mixing. 

While only a complete model spectrum can definitively predict the visibility of each molecule given a set of assumptions, we can use Figure \ref{Freedman} together with Figure
 \ref{columns} to make some generalizations.
 We defer the task of properly including our new chemical network into a complete model atmosphere to the future.

The absorption cross section per molecule for the CO band head at $2.3\,\rm \mu m$ is about $10^{-21}\,\rm cm^{2}$ per molecule. We know from model comparisons to data \citep[e.g.,][]{Cushing2008} that by $T_{\rm eff}=1100\,\rm K$ at $g=10^5$ this band head produces only a slight spectral feature.  Consulting Figure \ref{columns}
 and scaling appropriately we expect, all else being equal, that CO would remain
 detectable down to $T_{\rm eff}\sim 900\, \rm K$ at $g=10^4$ and 500 K at  $g=10^3$.
 This temperature is considerably cooler than expected by the conventional wisdom.
 
We likewise can consider the appearance of the $2.2\,\rm \mu m$ methane band which by definition first appears at the T0 spectral type
at $T_{\rm eff}\sim 1200\,\rm K$.
Figure \ref{columns} suggests that a comparable gravity-adjusted column abundance of methane will not appear in a 
Saturn-like $g = 10^3$ planet until $T_{\rm eff} \sim 650\,\rm K$. 
The directly imaged companion GJ 504 b \citep{Kuzu2013, Janson2013} provides a test of this reasoning.
The mass and effective temperature of GJ 504 b have been estimated at  $\sim5\,\rm M_J$ and 600 K. 
Its gravity should be $\sim\!6000\,\rm cm\,s^{-2}$. 
By the reasoning above methane would be expected and indeed has been detected \citep{Janson2013}.
 
Methane is far more detectable in the $3.3\,\rm \mu m$ fundamental band where the absorption cross section is over an order of magnitude larger.  $\rm CH_4$ has been 
detected here as early as spectral type L5 \citep{Noll2000} in field dwarfs. Likewise spectra over this wavelength range would better help to constrain the arrival of methane at low gravity in the directly imaged planets. Indeed \citet{Skemer2014} report a possible detection of this same feature in HR 8799 c, suggesting a trace abundance in the upper atmosphere.  

Another test of these ideas may be provided by field T or Y dwarfs with unusual colors not predicted by models. Faint, red objects in particular would be
good candidates for the low gravity, methane-poor objects we predict.  For example 
the cool Y dwarf WISE 1828+2650 \citep{Cushing2011}, is both faint and unusually red in $J-H$ color \citep{Beich2014}. 
While Cushing et al.\ did detect methane in this object, 
chemical equilibrium models do a very poor job fitting the available photometry \citep{Beich2014}.
Near-IR spectra should be obtained of all such color outliers to search for methane spectral features.

We note that some searches for  planetary companions to young stars \citep[e.g.,][]{Liu2010} as well as surveys for  `planetary-mass' T-dwarfs in young stellar clusters \citep[e.g.,][]{Parker2013} have employed 
differential-methane band imaging. In this technique \citep{Tinney2005} two images of a target are taken, one in a filter that matches the $1.7-\,\rm \mu m$ $\rm CH_4$ band and one 
which probes the entire {\it H} band. When the two images are differenced, methane-bearing objects stand out as they are dark in the $\rm CH_4$ filter. 
Our conclusions here suggest that such techniques must be used with caution as methane may simply not yet be present in planetary-mass objects even at effective temperatures below 1000 K.

Similar arguments can be made for the appearance of other spectral features of interest.
 We expect that $\rm NH_3$ will appear at effective temperatures about $200\,\rm K$ cooler in planets 
 compared to field brown dwarfs.
 HCN, with a cross section of $10^{-20}\,\rm cm^{2}$ per molecule at $1.55\,\rm \mu m$ is unlikely to be detectable in most objects as the computed column abundances are less than $10^{18}\,\rm cm^{-2}$.
HCN's prospects are poor at 3.0 microns despite a high cross-section unless the C/O ratio is higher than solar and water's abundance reduced.
Water's cross section is 100-fold smaller, but with solar abundances its column ($1-2\times 10^{21}$ cm$^{-2}$) is 1000 times what HCN can reach at its best.  
 Carbon dioxide has an absorption cross section $10^{-17}\,\rm cm^{2}$ per molecule at $4.2\,\mu\rm m$, and with abundances approaching $10^{18}\,\rm cm^{-2}$ we expect it to be detectable at around 900 to 1100 K in field brown dwarfs. Indeed the AKARI space telescope discovered $\rm CO_2$ features in several late L and early T dwarfs  \citep{Yam2010}. Judging by Figure \ref{columns}, we would predict $\rm CO_2$ to be detectable to 500 K and cooler in the lowest mass planets.

\section{Conclusions}

We use a reasonably complete 1-D chemical kinetics code to
survey the parameter space that encompasses
atmospheres of cool brown dwarfs and
warm young extrasolar giant planets.
Our model contains only gas phase chemistry of small molecules containing H, C, N, O, and S.
We use realistic $p$-$T$ profiles for cloudless atmospheres with effective temperatures between
500 and 1100 K and surface gravities between $10^3$ cm/s$^2$ to $10^5$ cm/s$^2$.
Vertical transport is described by an eddy diffusivity $K_{zz}$ that we vary over a wide range.
Our objective is to describe carbon and nitrogen speciation, especially at lower (planetary) surface gravities.
Overviews of what we found are presented in 
Figure \ref{ensemble} for carbon and 
Figure \ref{ZZ} for nitrogen. 

We find that carbon in cloudless brown dwarfs is predominantly in the form of methane at 900 K for $g=10^5$ cm/s$^2$. 
The small surface gravity of planets strongly discriminates against CH$_4$ when compared to an otherwise comparable brown dwarf.
 If vertical mixing is comparable to Jupiter's, methane first predominates over CO in planets cooler than 500 K.
 Sluggish vertical mixing can raise the transition to 600 K;
 clouds or more vigorous vertical mixing could lower it to 400 K.

The  detectability of specific molecular features in a spectrum depends on the strength of the molecular absorption cross sections as
well as the gaseous abundance. 
Nevertheless a natural prediction of our model is that there will be cool planets with no methane observed
in the H or K spectral bands.  
 The refractory behavior of CO in low gravity objects is likely at least partially responsible for the lack of cool planets  
 found by the NICI survey, which relied upon the methane absorption H band to identify planets \citep{Liu2010}.

Ammonia is also sensitive to gravity, 
but unlike methane and CO, ammonia is insensitive to mixing, which makes it a proxy for gravity.
We did not explore temperatures low enough to determine the transition from N$_2$ to ammonia
in planets, but it is nearly as abundant as N$_2$ 
at 500 K in brown dwarfs, which is broadly consistent with the 
observed properties of the Y dwarfs \citep{Cushing2011} for which $\rm NH_3$ is seen in H band.
On the other hand, ammonia persists as an abundant minor species to rather high temperatures
and this is
consistent with it being readily detected in mid-IR spectra of T-dwarfs \citep{Cushing2006}.
 HCN might become interesting in high gravity brown dwarfs if vertical mixing is very vigorous.

\medskip
When expressed in terms of quenching parameters, nearly all our results for CO and CH$_4$ can be reduced to a 
simple equation in Arrhenius form that is easy to use but not easy to interpret.
From a strictly practical perspective, what we find is close to what one gets
using \citet{Prinn1977}'s algorithm in its original form 
(a fortuitous accident of using too high a rate for the wrong reaction).
The apparent simplicity is somewhat surprising given the complexity of the chemistry
involved and the many different ways the system has been described in the literature.
Nor is our result quite what we expected.
First, we find that the timescale $t_{\rm CO}$ for hydrogenation of CO is shorter for higher metallicity. 
Because the forward reactions that destroy CO have no metallicity dependence,
we infer that metallicity dependence enters through the reverse reaction (oxidation of CH$_4$).
By definition the reverse reaction is as fast as the forward reaction while in equilibrium,
 but it falls off more quickly than the forward reaction as the temperature drops.
The important forward reactions are linear in metallicity whilst the important reverse reactions are quadratic in metallicity.

Second, we find that $t_{\rm CO}$ is inversely proportional to pressure.
Analyses based on isolating a limiting reaction predict a stronger dependence on pressure, typically $p^{-2}$.
The weaker $p^{-1}$ pressure dependence is expected for a reaction involving
CO itself, such as 
\begin{equation}
\label{H2+CO}
{\rm H}_2 + {\rm CO} \rightarrow {\rm HCO} + {\rm H}.
\end{equation}
It is interesting that R\ref{H2+CO} actually does have the highest $\Delta G$ value
in the CO hydrogenation sequence, and therefore the highest energy barrier, and that the magnitude of 
$\Delta G$ of R\ref{H2+CO} is what we infer in Eq \ref{t_chem}.
But it is also quite clear from Figure \ref{YY} that equilibrium breaks down between CH$_3$OH and CH$_4$
well before if breaks down between CO and CH$_3$OH,
as was found by \citet{Moses2011}.  
Reactions involving CH$_3$OH imply a $p^{-2}$ pressure dependence,
because the equilibrium abundance of CH$_3$OH with respect to CO goes as the square of the H$_2$ pressure. 
The system behaves---both in the 40,000 K energy barrier and the $p^{-1}$ pressure dependence---
 as if the overall rate depends on the entry reaction R\ref{H2+CO}
rather than on the rate of R\ref{methanol} in which equilibrium actually breaks down.
 The apparent paradoxes might be resolved through a more complete analysis of the 
reverse reactions, or they may result from systematic biases stemming from
our treating the mixing length as equal to a scale height,
 rather than as the complex function of many variables that it probably is \citep{Smith1998}.

For nitrogen and ammonia we were unable to recover an emergent Arrhenius-like behavior from the full system.
On the other hand, it was easy to devise a good quench approximation,
and we also found that some published quench approximations \citep[e.g.,][]{Fegley1994} work very well.  
For HCN, we met with mixed success.
We did find emergent Arrhenius-like behaviors, and
we developed a new quench approximation that works quite well for our models.
On the other hand there does not yet appear to be a consensus on what the quench chemistry of HCN actually is.
We found a wide scatter of different outcomes when comparing different published quench schemes.
We think our quench approximation should work well for warm objects where HCN is
predicted to be relatively abundant but, 
because we did not consider hypothetical hydrogenation channels through methylamine \citep{Moses2010}, our models
may overestimate $f_{\rm HCN}$, especially in cooler objects. 

\medskip
There are many effects that our models do not address.
None of these effects have much sway over our quench approximations, but they have much to do
with what might actually be present in a real atmosphere.

Clouds are likely the most important.
Adding infrared opacity to an atmosphere is effectively equivalent to reducing the gravity.
This is because opacity raises the $p$-$T$ profile (i.e., the adiabat) to a lower pressure for a given effective temperature.
A cloudy atmosphere favors CO and N$_2$ over CH$_4$ and NH$_3$, other things equal.
Increasing metallicity also increases opacity to the detriment of methane and ammonia.

Another issue for real worlds is that $K_{zz}$ may be much smaller in higher altitude radiative regions of the atmosphere
 than it is in the convecting region. 
The radiative stratosphere is the part of the atmosphere where $K_{zz}$ the modeling parameter is most suspect.
For N$_2$ and NH$_3$ pronounced vertical structure in $K_{zz}$ will not matter much.
For CO and CH$_4$, sharply lower values of $K_{zz}$ in the stratosphere might be important, because 
if $K_{zz}$ is low enough, the atmosphere can have a second quench point in the stratosphere.
Such behavior no problem to a full kinetics model, but it can make implementing 
quench approximations more complicated.

Photochemistry induced by irradiation from a nearby star can deplete NH$_3$ and increase CO$_2$ and HCN.
Lightning and/or impact shocks in cool NH$_3$- and CH$_4$-rich atmospheres can generate CO and HCN \citep{Chameides1981}. 
In warmer objects, catalysts that may be present in metallic clouds 
(here we mean real metals) 
will lower the quench point and thus will favor NH$_3$, CH$_4$, and CO$_4$ \citep{Prinn1987}.

All of these ideas will be put to the test by the directly imaged planets that are expected to be discovered in the coming months and years. Constraining 
the composition of their atmospheres will no doubt be a rewarding endeavour.

\section*{Acknowledgments}
The authors thank Richard Freedman, Caroline Morley,  
Julianne Moses, Didier Saumon, and Channon Visscher for many insightful discussions and occasional course corrections.
The authors thank the NASA Planetary Atmospheres Program for support of this work.

\end{document}